\documentclass[aps,prd,nofootinbib,twocolumn,floatfix,eqsecnum,showpacs]{revtex4}
\usepackage{color}
\usepackage{amssymb}
\usepackage{graphicx}
\newcommand{\tr}{\mbox{Tr}\;}
\newcommand{\M}{\mathbb{D}}
\newcommand{\Olejnik}{\v{S}.\ Olejn{\'\i}k}
\newcommand{\gtwo}{G${}_2$\ }
\begin{document}
%
\title{Casimir scaling in G${}_2$ lattice gauge theory}
\author{{L\kern-.08cm'}udov{\'\i}t Lipt\'ak}
\email{ludovit.liptak@savba.sk}
\author{{\v{S}}tefan Olejn{\'\i}k}
\email{stefan.olejnik@savba.sk}
\affiliation{Institute of Physics, Slovak Academy of Sciences, SK--845 11 Bratislava, Slovakia}
\date{\today}
\begin{abstract}
We computed potentials between static color sources from the six lowest representations of \gtwo lattice gauge theory, in numerical simulations with the Wilson action on asymmetric lattices with nonperturbatively estimated values of the bare anisotropy. We present evidence for (approximate) Casimir scaling of the obtained intermediate string tensions. The agreement with the Casimir-scaling prediction improves by increasing the coupling $\beta$ in the weak-coupling region above the crossover observed in \gtwo gauge theory. The result naturally fits into confinement models with magnetic disorder and vacuum domain structure, but may represent a challenge for other approaches.
\end{abstract}
\pacs{11.15.Ha, 12.38.Aw}
\maketitle
\section{Introduction}\label{sec:intro}
	Any theory aiming to explain the mechanism of color confinement in quantum chromodynamics has to face a would-be simple -- but rather non-trivial in reality --  task: to describe qualitative features of potentials between static charges from various representations of the gauge group. If one neglects dynamical quarks in the first approximation, the static quark-antiquark potential in SU($N$) gauge theory exhibits distinct behavior in three ranges of interquark distances. At short distances, the interaction is dominated by gluon exchange and the potential is Coulomb-like, its strength being proportional to the eigenvalue of the quadratic Casimir operator in the given representation of color sources. At intermediate distance scales, from the onset of confinement to the onset of color screening, the potential is expected to be linearly rising, and the corresponding string tensions for different representations are again approximately proportional to the quadratic Casimir.\footnote{This proportionality is exact in two dimensions, and can be extended to $D=3$ and 4 by dimensional reduction arguments, see \textit{e.g.\/}\ \cite{Greensite:1979yn,Olesen:1981zp,Ambjorn:1984mb}.} This effect, dubbed Casimir scaling in Ref.~\cite{DelDebbio:1995gc}, was observed in numerical simulations of both SU(2) \cite{Ambjorn:1984mb,Michael:1985ne,Piccioni:2005un} and SU(3) \cite{Markum:1988na,Campbell:1985kp,Deldar:1999vi,Bali:2000un,Gupta:2007ax} lattice gauge theories.\footnote{Ref.\ \cite{Bali:2000un} contains an extensive bibliography of earlier references.} Finally, at asymptotic distances, the string tensions depend only on the $N$-ality $k$ of the representation: representations with $k=0$ are screened and their potentials are asymptotically flat, $k\ne 0$ representations are confined and their potentials continue to rise linearly.

	A confinement scenario which naturally accounts for both Casimir scaling at intermediate distances and the asymptotic $N$-ality dependence is based on condensation of center vortices in the QCD vacuum (see \cite{Greensite:2003bk} for a review and summary of numerical evidence for the model). The vacuum is filled with percolating thick vortices carrying color magnetic flux quantized in terms of elements of the gauge group center. Asymptotically large Wilson loops, linked to center vortices, ``feel'' the total (center) flux; this is how $N$-ality shows up easily in asymptotic string tensions. On the other hand, Wilson loops of intermediate size cross thick-vortex cores and enclose only a part of their magnetic flux. As a consequence, intermediate string tensions can exhibit approximate Casimir scaling. A simple model of how and why it happens, was suggested in Ref.~\cite{Faber:1997rp} (some numerical support was provided by \cite{Faber:1998qn}).
	
	Recently, \gtwo Yang--Mills theory has attracted considerable attention \cite{Holland:2003jy,Pepe:2005sz,Pepe:2006er,Cossu:2006bi,Cossu:2007dk,Cossu:2007ns,Greensite:2006sm,Maas:2007af} as a means to look at the physics of confinement from a fresh new angle. The \gtwo group has a trivial center; the center of its universal covering group is also trivial\cite{Holland:2003jy}, and there seems to be no reason to believe that the group center plays any important role in \gtwo gauge theory at all. In the strict sense of the word the theory is only \textit{temporarily\/} confining, since potentials for static color charges from any representation, including the fundamental one, must be asymptotically flat. Unlike SU($N$), even a ``quark'' from the fundamental (\{7\}) representation of \gtwo can be screened by a bunch of ``gluons'' (from the 14-dimensional adjoint representation). However, this fact does not really contradict the center-vortex confinement scenario: \gtwo does not possess non-trivial center vortices, and the asymptotic string tension is zero if they are not present. As argued by the authors of Ref.\ \cite{Greensite:2006sm}, the \gtwo temporary vs.\ permanent confinement issue is more semantics than physics; the really important point is that even in the \gtwo gauge theory one expects the static potentials grow linearly over a certain range of distances, from the scale where perturbation theory breaks to the onset of screening. The linear rise of the fundamental potential in this intermediate region was clearly demonstrated in numerical simulations \cite{Greensite:2006sm}.

	All asymptotic string tensions are zero in \gtwo -- what about Casimir scaling at intermediate distances? Dimensional reduction could be invoked to argue in its favor. However, if it holds true and if finite thickness of center vortices was able to explain successfully Casimir scaling in SU($N$), what can it be attributed to in a theory with trivial center? Is there any origin of Casimir scaling common for ``centerless'' and ``centerful'' gauge models? The answer given in \cite{Greensite:2006sm} was affirmative, the authors provided a model -- an extension and improvement of that of Ref.~\cite{Faber:1997rp} -- in which the Yang--Mills vacuum state has a domain structure, with the color magnetic flux in each domain quantized in units of the gauge group center (be it trivial or not), and Casimir scaling results from random spatial fluctuations of the flux in each domain. 
	
	The aim of the present paper is to subject the Casimir-scaling hypothesis and predictions of the model of \cite{Greensite:2006sm} to a test in numerical simulations of \gtwo gauge theory on a four-dimensional Euclidean lattice. First, in Sec.\ \ref{sec:model}, we will briefly summarize elements of the model. Then, in Sec.\ \ref{sec:preliminaries}, we sum up a few facts about the exceptional \gtwo group, explain the applied lattice methods, and present a subset of technical details on the enhancement of the ground-state overlap via smearing, and the determination of renormalized anisotropies. Section \ref{sec:potentials} summarizes results, in particular subsection \ref{subsec:fundamental} the fundamental representation potential, while \ref{subsec:higher} potentials for higher representations. We conclude with a brief discussion and summary (Sec.\ \ref{sec:conclusions}). A few technicalities are relegated to appendices. 

\section{A model of the Yang--Mills vacuum: Recapitulation}\label{sec:model}
	An essential element of the derivation of Casimir scaling in the $D=2$ Yang--Mills theory is the following: the YM action is quadratic in the field strength tensor $F_{ij}$, contains no derivatives of or constraints on $F_{ij}$, therefore field strengths of vacuum fluctuations in different points are uncorrelated, and disorder all-representation Wilson loops. This cannot be true for an effective action in $D>2$ dimensions. Even though the leading term of a derivative expansion of the Yang--Mills effective action is again quadratic in field strengths \cite{Greensite:1979yn},\footnote{See recent arguments in \cite{Greensite:2007ij} and references therein.} the fields are constrained by Bianchi identities and higher-order corrections contain derivatives. This is not a drawback, but desired: constraints and derivative terms must somehow contrive, in $D>2$ dimensions, to change the proportionality of string tensions to quadratic Casimirs to the asymptotic $N$-ality dependence.
	
	In the simple model of Ref.\ \cite{Greensite:2006sm} it is assumed that if we take a 2D slice of the four-dimensional volume, we can split it into (partly) overlapping domains (``patches'') of a typi\-cal area $A_d$. Within each domain color magnetic fields fluctuate randomly and (almost) independently, with a short length of correlation $l$. Each domain is a bunch of small independently fluctuating subregions of area $l^2\ll A_d$. The only weak constraint on fluctuating fields bounds the total integrated magnetic flux over each domain to correspond to an element of the center of the gauge group. In SU(2), there are two types of domains, of the center-vortex type and of the vacuum type. The former correspond to the nontrivial element of the center, $-\mathbb{I}$, and represent a cross section of a thick center vortex; the latter carry a zero total magnetic flux. In \gtwo all domains will be of the vacuum type, since the center contains the identity element only. How to distinguish overlapping vacuum domains in 2D snapshots of the \gtwo vacuum remains an unresolved task for future reflection and/or numerical experiments.  
	
	Without going into details, let's briefly summarize conclusions of the model. If the center of the gauge group contains $N$ elements, there are $N$ types of vacuum domains enumerated by the value $k\in \{0, 1, \dots, N-1\}$, each assumed to appear with a probability $f_k$ centered at any given plaquette in the plane of the loop. The effect of a domain (a 2D cross section of the $k$th vortex) on a planar Wilson loop is to multiply the loop by a group element
\begin{equation}
\mathcal{G}(\alpha^{(k)},\mathcal{S})=\mathcal{S}\;\exp\left[i\vec{\alpha}^{(k)}\cdot\vec{\mathcal{H}}\right]\; \mathcal{S}^\dagger,
\end{equation}
where $\{\mathcal{H}_i\}$ are generators of the Cartan subalgebra, $\mathcal{S}$ is a random element of the group, 
and angles $\vec{\alpha}^{(k)}$ depend on the location of the vortex/domain with respect to the loop. If the domain is all contained within the loop area, then
\begin{equation}
\exp\left[i\vec{\alpha}^{(k)}\cdot\vec{\mathcal{H}}\right]=z_k\mathbb{I},
\end{equation}
where $z_k$ is the $k$th element of the center, $\mathbb{I}$ is the unit element. If the domain is outside the loop, it has no effect on the Wilson loop, \textit{i.e.\/}
\begin{equation}
\exp\left[i\vec{\alpha}^{(k)}\cdot\vec{\mathcal{H}}\right]=\mathbb{I}.
\end{equation}
For a Wilson loop from the representation $r$, the averaged contribution of a domain is simply
\begin{equation}
\mathcal{G}_r(\alpha^{(k)})\; \mathbb{I}_{d_r}=\frac{1}{d_r}\chi_r\left(\exp\left[i\vec{\alpha}^{(k)}\cdot\vec{\mathcal{H}}\right]\right)\; \mathbb{I}_{d_r},
\end{equation}
where $d_r$ is the dimension of the representation $r$ and $\mathbb{I}_{d_r}$ is the $d_r\times d_r$ unit matrix.

	One further assumes that probabilities to find domains of any type centered at two different plaquettes are independent, and that for loops smaller than the typi\-cal size of the domain, the r.m.s.\ of phases $\alpha$ is proportional to the area of the vortex contained in the interior of the loop. Then it is a simple exercise to show that, both in SU($N$) and G${}_2$, the static potential $V_r(R)$ of the representation $r$ will be linearly rising for distances $l\ll R\ll\sqrt{A_d}$, with a string tension approximately proportional to its quadratic Casimir:
\begin{equation}
\sigma_r\sim C_r\qquad\mbox{(intermediate distances).}
\end{equation}
For very large Wilson loops most vortices will be contained within the loop, the average phases are proportional to the corresponding total magnetic flux through the domain, and the prediction is, for $R\gg\sqrt{A_d}$
\begin{equation}
\sigma_r\sim F[k_r, \{f_k\}]\qquad\mbox{(asymptotic distances)}
\end{equation}
for SU($N$), where $k_r$ is the $N$-ality of the representation~$r$, and 
\begin{equation}
\sigma_r=0\qquad\mbox{(asymptotic distances)}
\end{equation}
for the \gtwo gauge group. The function $F$ above depends also on the set of probabilities $\{f_k\}$; these can be tuned to get a desired form of $k$-scaling of asymptotic string tensions (see \cite{Greensite:2002yn} for a discussion of this point).

	Thus, the domain model of the YM vacuum with magnetic disorder \cite{Greensite:2006sm}, containing only a handful of adjustable parameters, predicts Casimir scaling of higher representation potentials at intermediate distances \textit{both\/} for SU($N$) and (centerless) \gtwo gauge theory. To verify the prediction for G${}_2$, numerical simulations of the theory in lattice formulation are the appropriate tool. 

\section{\gtwo on a lattice: Preliminaries}\label{sec:preliminaries}

\subsection{\gtwo group}\label{subsec:g2}

\begin{table}
\caption{Group theoretical factors for G${}_2$. $\{ D \}$ is the representation, $D$ its dimension, $ [\lambda_1,\lambda_2] $ are its Dynkin coefficients, and $ d_{\{D\}} = C_{\{D\}} / C_F $ denotes the ratio of its quadratic Casimir to that of the fundamental representation.}\label{tab:casimirs}
\begin{ruledtabular}
\begin{tabular}{cccccccc}
$\{D\}$           &  $ \{7\} $     &  $ \{14\} $     &  $ \{27\} $     &  $ \{64\} $     &  $ \{77\} $     &  $ \{ 77'\} $   \\
$[\lambda_1,\lambda_2]$   & $ [1,0] $  & $ [0,1] $ & $ [2,0] $ & $ [1,1] $ & $ [3,0] $ & $ [0,2] $   \\
$ d_{\{D\}} $  & $ 1 $      & $ 2 $     & $ 2.\bar{3} $     & $ 3.5 $     & $ 4 $     & $ 5 $     \\
\end{tabular}
\end{ruledtabular}
\end{table}

	\gtwo is the smallest of the 5 exceptional simple Lie groups, at the same time the simplest for which the universal covering group is the group itself, and both have a trivial center~\cite{Holland:2003jy}. The basic information on the group can be found \textit{e.g.\/}\ in \cite{Holland:2003jy,Pepe:2006er}, a part is summarized here for the sake of completeness. The \gtwo group has rank 2 and 14 generators. Its fundamental representation is 7-dimensional. \gtwo is real and a subgroup of SO(7) of rank 3 with 21 generators, a group of $7\times 7$ orthogonal matrices $\M$ with unit determinant. \gtwo matrices satisfy 7 additional cubic constraints:
\begin{eqnarray}
T_{abc} &=& T_{def}\M_{da}\M_{
eb}\M_{
fc},\\
\nonumber
T_{127} = T_{154}&=&T_{163}= T_{235} = T_{264} = T_{374} = T_{576} = 1,
\end{eqnarray}
where $T$ is an antisymmetric tensor non-zero elements of which are listed above (up to permutations of indices).

	For a study of higher representations we will need tensor decompositions of various tensor products. Group theory provides the following relations:\footnote{The decompositions were obtained using {\sffamily\bfseries LiE}, a clever computer algebra package for Lie group computations. For details, see \cite{vanLeeuwen:1994aa}.}
\begin{eqnarray}\label{eq:decomp}
 \{7\}  \otimes  \{7\}  & = & \{1\}  \oplus  \{7\}  \oplus  \{14\}  \oplus  \{27\}, \nonumber \\
 \{7\}  \otimes  \{14\} &  = & \{7\}  \oplus  \{27\}  \oplus  \{64\}, \\
 \{14\}  \otimes  \{14\} &  = &  \{1\}  \oplus  \{14\}  \oplus  \{27\}  \oplus  \{77\}  \oplus \{77'\},   \nonumber \\
 \{7\}  \otimes  \{27\} &  = &  \{7\}  \oplus  \{14\}  \oplus  \{27\}  \oplus  \{64\}  \oplus \{77\}. \nonumber 
\end{eqnarray}

	Assume now $\M^{\{D\}}(g)$ is a matrix corresponding to an element $g\in{}$\gtwo in its $D$-dimensional irreducible representation. Then the relations (\ref{eq:decomp}) imply that traces of higher-representation matrices can be expressed through traces of the fundamental- and adjoint-representation matrices:\footnote{We suppressed the argument $g$ and used $F$ and $A$ instead of $\{7\}$ and $\{14\}$ to denote the fundamental and adjoint representation respectively. The symbol Tr is understood in the usual matrix sense, \textit{i.e.\/} $\tr\M^{\{d\}}=\sum_{a=1}^{d}\M^{\{d\}}_{aa}$.}${}^,$\footnote{A careful reader might notice that, in the case of SU(3), a well-known decomposition $\{3\}\otimes\{3\}=\{6\}\oplus\{\bar{3}\}$ 
leads to $\tr\M^{\{6\}}=(\tr\M^F)^2-(\tr\M^F)^*$, while Refs.\ \cite{Deldar:1999vi,Bali:2000un} used $\tr\M^{\{6\}}=\frac{1}{2}\left[(\tr\M^F)^2+\tr(\M^F)^2\right]$. Both formulas (and similar pairs for other representations of SU(3)) can be shown equivalent using the Cayley--Hamilton theorem.}
\begin{eqnarray}\label{eq:trace_formulas}
&& {\tr} \M^{\{27\}} = - 1 - {\tr} \M^A - {\tr} \M^F +  ({\tr} \M^F)^2, \nonumber \\
&& {\tr} \M^{\{64\}} = 1  +  {\tr} \M^A +{\tr} \M^A\cdot {\tr} \M^F -  ({\tr} \M^F)^2, \nonumber \\
&& {\tr} \M^{\{77\}} = - {\tr} \M^A - {\tr} \M^F - 2\;{\tr} \M^A \cdot {\tr} \M^F \nonumber\\
&& \phantom{{\tr} \M^{\{77\}} = } - ({\tr} \M^F)^2 + ({\tr} \M^F)^3, \nonumber \\
&& {\tr} \M^{\{77 ' \}} = {\tr} \M^A + ({\tr} \M^A)^2 + 2\;{\tr} \M^F  \nonumber\\
&& \phantom{{\tr} \M^{\{77 ' \}} = } + 2\;{\tr} \M^A\cdot {\tr} \M^F - ({\tr} \M^F)^3. 
\end{eqnarray}
The adjoint-representation matrix is constructed from the fundamental one:
\begin{equation}\label{eq:fun_to_adj}
\M^A_{ab}(g) = 2\;\tr\left[ \M^F(g)^\dagger \; t_a \; \M^F(g) \; t_b \right]. 
\end{equation}

	The eigenvalue of the quadratic Casimir operator for the $D$-dimensional irreducible representation $\{D\}$ of the \gtwo group can be computed from its Dynkin coefficients $[\lambda_1,\lambda_2]$ (see \textit{e.g.}\ \cite{Cahn:1985wk}; we adopted the convention used in~\cite{vanLeeuwen:1994aa}). The dimension $D$ of a representation of \gtwo with Dynkin coefficients $[\lambda_1,\lambda_2]$ is given by \cite{Engelfield:1980ab}:
\begin{equation}
D=\frac{9}{40}(\ell_1^2 - \ell_2^2)(\ell_2^2 - \ell_3^2)(\ell_3^2 - \ell_1^2),
\end{equation}
where $\ell_1 = \frac{1}{3} (1+\lambda_1)$, $\ell_2 = \frac{1}{3} (4 + \lambda_1 + 3\lambda_2)$, 
and $\ell_3 = \frac{1}{3} (5+2\lambda_1 + 3\lambda_2)$, while the ratio of its quadratic Casimir operator to that of the fundamental representation is~\cite{Engelfield:1980ab}:
\begin{equation}\label{eq:casimir}
d_{\{D\}}\equiv\frac{C_{\{D\}}}{C_F}=\frac{1}{4}\left(\ell_1^2 + \ell_2^2 + \ell_3^2 - \frac{14}{3}\right).
\end{equation}
Explicit values for the six lowest representations of \gtwo are listed in Table~\ref{tab:casimirs}.

\subsection{\gtwo lattice gauge theory}\label{subsec:g2_on_lattice}

	We closely follow the lattice formulation of \gtwo gauge theory outlined in \cite{Pepe:2005sz}. On links of a space-time lattice, one puts matrices $U_\mu(x)$ from the fundamental ($\{7\}$) representation of G${}_2$. Those are represented by $7\times 7$ complex matrices in a parametrization suggested by \cite{Macfarlane:2002hr}, which makes explicit the separation into the G${}_2$/SU(3) coset group and the SU(3) subgroup.\footnote{Another parametrization of \gtwo matrices, via Euler angles \cite{Cacciatori:2005yb} using real matrices, was described in Appendix B of Ref.\ \cite{ Greensite:2006sm}.} (Details of the parametrization can be found in Pepe's review \cite{Pepe:2005sz}.)  Generators of the \gtwo algebra corresponding to this parametrization are listed in Appendix \ref{sec:generators}.\footnote{We are grateful to Axel Maas for deriving explicit expressions.}

	We use the Wilson action:
\begin{eqnarray}\label{eq:Wilson_action}
S &=& - \frac{\beta}{7} \sum_{x, i > 0} \xi_0 \; \textrm{Re Tr } \left[   P_{i0} (x)  \right] \nonumber\\
&& - \frac{\beta}{7} \sum_{x, i > j > 0} \frac{1}{\xi_0} \textrm{Re Tr } \left[   P_{ij} (x)  \right], 
\end{eqnarray} 
where $ P_{\nu \tau} (x) = U_{\nu} (x) \; U_{\tau} (x + \hat{\nu}) \; U^{\dagger}_{\nu} (x + \hat{\tau}) \; U^{\dagger}_{\tau} (x)  $. Thermalized lattice configurations can be generated by a combination of pure-SU(3) Cabibbo--Marinari updates followed by random \gtwo transformations,
and overrelaxation sweeps, see \cite{Maas:2007af} for a more detailed description. 

	In most cases we work on asymmetric lattices with different lattice spacings in time ($a_t$) and space ($a_s$) directions, therefore the action (\ref{eq:Wilson_action}) contains a parameter $\xi_0$ called the \textit{bare anisotropy}. To reproduce the usual classical gauge action in the naive continuum limit, the bare anisotropy has to be chosen as $\xi_0=a_s/a_t$. However, what we need to achieve is a certain fixed ratio of
lattice spacings in space and time directions \textit{in physical units}, $\xi=a_s^{\mathrm{phys}}/a_t^{\mathrm{phys}}$.
The \textit{renormalized anisotropy} $\xi$ is a function of $\xi_0$ and $\beta$ in the Wilson action (\ref{eq:Wilson_action}), below we will discuss a nonperturbative determination of $\xi_0=f(\xi,\beta)$ to get a desired value of $\xi$ for a given coupling $\beta$. Before that, we will recall a few basic facts about the determination of the static potential from rectangular Wilson loops.
	
\subsection{Static potentials from the lattice}\label{subsec:asym_potentials}

	Let us start from a symmetric lattice with lattice spacing $a$. On a set of thermalized lattice configurations, we measure expectation values $W(r,t)$ of Wilson loops of spatial extent $\mathbf{r}=r\cdot a$ and a temporal separation $\mathbf{t}=t\cdot a$.\footnote{Boldface symbols, $\mathbf{x}, \mathbf{t}, \dots$, denote variables with the dimension of length/time, $x, t, \dots$, are used for dimensionless quantities, \textit{i.e.\/}\ $\mathbf{x}, \mathbf{t}, \dots$, divided by $a$, or $a_s, a_t$ on an asymmetric lattice.} For large time separations
\begin{equation}\label{eq:W_at_large_t}
W(r,t)=c(\mathbf{r},a)\exp[-V(\mathbf{r},a)t]\qquad\mbox{for large }t.
\end{equation}
The static potential
\begin{equation}
V(\mathbf{r},a)=-\lim_{\mathbf{t}\to\infty}\frac{1}{\mathbf{t}}\ln W(r,t)
\end{equation}
consists of two contributions: the true $\mathbf{r}$-dependent static potential due to the interaction of a color source and anti-source, and the self energy contribution (independent of~$\mathbf{r}$) divergent in the continuum limit. Thus we can write
\begin{equation}
V(\mathbf{r},a) = \frac{1}{a}\left[\widehat{V}^{\mathrm{int}}(\mathbf{r},a)+\widehat{V}^{\mathrm{self}}(\mathbf{r},a)\right].
\end{equation}
(Potentials with a hat are dimensionless.)
	
	On an asymmetric lattice one can introduce three different potentials, defined as:
\begin{eqnarray}
\label{eq:potss}
{V}_{ss} ( \mathbf{r},a_s )\Big|_{\mathbf{r}=\mathbf{x}} & = & -  \lim_{{y} \to \infty} \frac{1}{\textbf{y}}  \ln W_{ss} (x, y) \nonumber\\
& = & - \frac{1}{a_s}  \lim_{{y} \to \infty} \frac{1}{{y}}  \ln W_{ss} (x, y), \\
\label{eq:potst}
{V}_{st} ( \textbf{r} ,a_s)\Big|_{\mathbf{r}=\mathbf{x}} & = & -  \lim_{{t} \to \infty} \frac{1}{\textbf{t}}  \ln W_{st} (x, t) \nonumber\\
& = & -  \frac{1}{a_t} \lim_{{t} \to \infty} \frac{1}{{t}}  \ln W_{st} (x, t), \\
\label{eq:potts}
{V}_{ts} ( \textbf{r},a_t)\Big|_{\mathbf{r}=\mathbf{t}} & = & -  \lim_{{y} \to \infty} \frac{1}{\textbf{y}}  \ln W_{st} (y,t) \nonumber \\
& = & -  \frac{1}{a_s} \lim_{{y} \to \infty} \frac{1}{{y}}  \ln W_{st} (y,t).
\end{eqnarray}
The Wilson loop $W_{ss}$ has both sides in spatial directions, while the second side of $W_{st}$ lies in the temporal direction (\textit{i.e.\/}\ the one in which the lattice is discretized more densely).

	One can argue for a simple relation between the interaction potentials \cite{Bali:2000un,Smit:2002ug}:
\begin{eqnarray}
\label{eq:pot_rel1}
\widehat{V}^{\mathrm{int}}_{ss} (\textbf{r}, a_s) &=& \xi\; \widehat{V}^{\mathrm{int}}_{st} (\textbf{r}, a_s) + \Delta  \widehat{V}^{\mathrm{self}},\\
\label{eq:pot_rel2}
\widehat{V}^{\mathrm{int}}_{ss} (\textbf{r}, a_s) &=&  \widehat{V}^{\mathrm{int}}_{ts} ( \xi \textbf{r}, a_t). 
\end{eqnarray}

\subsection{Renormalization of the anisotropy}\label{subsec:anisotropy}

\begin{table}[!t]
\caption{The bare anisotropy $\xi_0(r,y)$ obtained from different $(r,y)$ pairs entering Eq.\ (\protect\ref{eq:R_equation}).} \label{tab:xi_0}
\begin{ruledtabular}
\begin{tabular}{c c c c c}
 $\beta $ &  $\xi_0 (2,2)$   &  $\xi_0 (2,3)$  &  $\xi_0 (3,2)$  &  $\xi_0 (3,3)$      \\\hline
 9.5   & $1.572 (14)$ & $1.586 (45)$ & $1.580 (41)$  & $1.596 (279)$  \\
 9.7   & $1.626 (22)$ & $1.638 (47)$ & $1.627 (50)$ & $1.631 (155)$  \\
\end{tabular}
\end{ruledtabular}
\end{table}

	Arguments for the usefulness of working on an asymmetric lattice have been clearly discussed \textit{e.g.\/}\ in \cite{Bali:2000un,Klassen:1998ua}: a reduction of lattice artifacts, a certain hope to decrease errors due to the influence of the self energy $\widehat{V}^{\mathrm{self}}$, etc. The determination of static potentials at a fixed separation of sources relies on  relations like (\ref{eq:W_at_large_t}) and requires to evaluate $W$ at as many $t$ values as possible. Our main motivation for using asymmetric lattices thus was to have more data points available in the physical $t$ window. We worked with lattices $L^3\times(2L)$, but wanted to keep the total physical length of the lattice in all directions the same, \textit{i.e.\/}\ the renormalized anisotropy $\xi=a_s^{\mathrm{phys}}/a_t^{\mathrm{phys}}=2$. To determine the bare anisotropies $\xi_0=f(\xi,\beta)$ that lead to $\xi=2$ at the used values of the coupling $\beta$ we applied the procedure of \cite{Klassen:1998ua}. One requires that the static potentials, which one can introduce on an asymmetric lattice, fulfill conditions (\ref{eq:pot_rel1}) and (\ref{eq:pot_rel2}). To avoid the nontrivial task of reliable determination of  $\Delta  \widehat{V}^{\mathrm{self}}$ in Eq.\ (\ref{eq:pot_rel1}), we estimated $\xi_0$ using the latter condition.

	In practice, we computed, for a given $\beta$ and lattice size, quantities:
\begin{equation}
R_{ss}(r,y)=\frac{W_{ss}(r,y)}{W_{ss}(r+1,y)},\quad R_{st}(r,t)=\frac{W_{st}(r,t)}{W_{st}(r+1,t)},
\end{equation}
and searched for a value of $\xi_0$ to fulfill
\begin{equation}\label{eq:R_equation}
R_{ss}(r,y)=R_{st}(r,t=2 y).
\end{equation}
Of course, the obtained value of $\xi_0$ somewhat depends on the chosen pair of side lengths $(r,y)$ in Eq.\ (\ref{eq:R_equation}). Since larger Wilson loops have smaller values and larger errors, we restricted ourselves to combinations $(2 \div 3,2 \div 3)$ only.

	The described nonperturbative determination of $\xi_0$ was performed in simulations for $\beta=9.5$ (about 3000 configurations at each trial $\xi_0$) and $9.7$ (1000 configurations at each $\xi_0$) on a $12^3\times 24$ lattice. The results are summarized in Table \ref{tab:xi_0}. For production runs we chose values close to $\xi_0(2,3)$, but checked a rather weak dependence of results on its precise value. The value chosen for $\beta=9.6$ was obtained by linear interpolation.
	
	Our final set of bare anisotropies used in production runs was $\xi_0=1.590, 1.615, 1.640$, for $\beta=9.5, 9.6, 9.7$, respectively.

\subsection{Smearing}\label{subsec:smearing}

	Another problem of the determination of static potentials from Wilson loops is the insufficient overlap of the trial quark-antiquark state with the ground state, especially at larger spatial distances. An accepted remedy for this disease is not to construct Wilson loops directly from link matrices, but to use smeared links instead. Various variants of smearing have been proposed in the literature, we used the \textit{stout smearing procedure} of Morningstar and Peardon \cite{Morningstar:2003gk}. The smearing is performed in several steps: First we calculate a weighted sum $S_{\mu}(x)$ of staples neighboring the link $U_{\mu}(x)$:
\begin{eqnarray}
S_{\mu} (x) & = & \rho \sum_{\nu \neq \mu}  \Bigl[ U_{\nu} (x) U_{\mu} (x + \hat{\nu}) U^{\dagger}_{\nu} (x + \hat{\mu}) \nonumber\\  
& + &  U^{\dagger}_{- \nu} (x) U_{\mu} (x - \hat{\nu}) U_{\nu} (x - \hat{\nu} + \hat{\mu})\Bigr], 
\end{eqnarray}
$\rho$ is a free parameter of smearing. We sum over directions $\nu$ depending on the potential studied (perpendicular to the Wilson loop, used for calculation of the potential). In the second step we project the product 
$\Omega_{\mu}(x)= S_{\mu} (x) U_{\mu}^{\dagger} (x)$ (no sum over $\mu$!) into the \gtwo algebra using:
\begin{equation}
Q_{\mu} (x)  =  \sum_{a}  \mbox{Im} \left( \tr\left[ \Omega_{\mu}(x)\;t_a  \right]\right)  t_a, 
\end{equation} 
where $t_a$ are generators of the algebra (see Appendix \ref{sec:generators}). In the last step, the smeared link is calculated as
\begin{equation}
U_\mu^{\mathrm{sm}} (x) =  \exp \left[  i  Q_{\mu} (x)   \right]   U_{\mu} (x).
\end{equation}
The exponential function of matrices was calculated by means of spectral representation and LAPACK subroutines. 

	After an extensive exploration of the effect of the free parameter $\rho$ and of the number of smearing steps $N_\mathrm{sm}$ used on the ground-state overlap, we chose $\rho = 0.1$. The optimal number of smearing steps was determined for each kind of potential: $N_\mathrm{sm} = 20$ for $V_{st}$,  $N_\mathrm{sm} = 15$ for $V_{ss}$, and $N_\mathrm{sm} = 5$ for $V_{ts}$ (see Appendix~\ref{sec:overlap} for details). In production runs we always applied all three smearing procedures on each thermalized configuration separately, \textit{i.e.\/} each time we smeared a new configuration using different staples and different number of smearing steps to get best overlap for a given potential.

	As expected, smearing did enhance the ground-state overlaps considerably for all studied types of potentials. Not to overcrowd the main body of the paper with technicalities, some illustrative examples are supplied in Appendix~\ref{sec:overlap}.
	
\section{\gtwo on a lattice: Potentials}\label{sec:potentials}

\subsection{Determination of potentials and setting the physical scale}\label{subsec:fits}

\begin{table}[!t]
\caption{Simulation parameters. (Smearing parameters were listed in Sec.~\protect\ref{subsec:smearing}. In runs on symmetric lattices thermalized configurations were separated by 20 sweeps consisting of 5 updates described in Sec.~\protect\ref{subsec:g2_on_lattice} and 1 overrelaxation step, on asymmetric lattices we used configurations separated by 15 such sweeps.)} \label{tab:runs}
\begin{ruledtabular}
\begin{tabular}{c c c c}
$\beta$ & $\xi_0$ & lattice size & $N_\mathrm{conf}$ \\\hline
\multicolumn{4}{c}{\textit{--- symmetric lattices ---}}\\\hline
9.5 & 1 & $14^4$ & 400\\
9.6 & 1 & $14^4$ & 400\\
9.7 & 1 & $14^4$ & 400\\\hline
\multicolumn{4}{c}{\textit{--- asymmetric lattices ---}}\\\hline
9.5 & 1.590 & $14^3\times 28$ & 960\\
9.6 & 1.615 & $14^3\times 28$ & 940\\
9.7 & 1.640 & $14^3\times 28$ & 960
\end{tabular}
\end{ruledtabular}
\end{table}
\begin{figure}[!tbhp]
\includegraphics[width=8truecm]{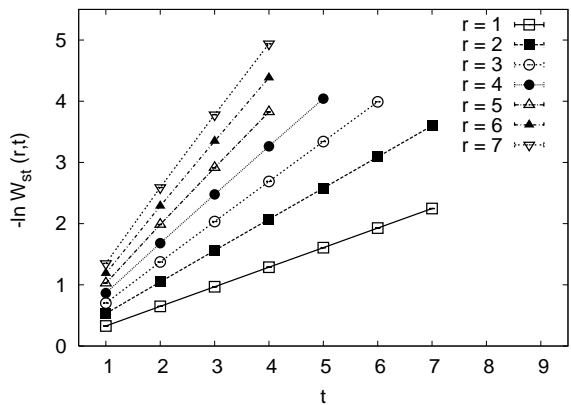} 
\caption{Logarithm of the expectation value of the fundamental Wilson loop $W_{st}(r,t)$ for fixed $r$-values as a function of $t$. ($14^3 \times 28$ lattice, $\beta = 9.6$. Loops were computed from smeared links. Lines are drawn to guide the eye.)}
\label{fig:potbf14_96_st}
\end{figure}

	 As discussed earlier (Sec.\ \ref{subsec:asym_potentials}), on an asymmetric lattice one can determine three kinds of potentials from the expectation values of Wilson loops of various orientations, $W_\lambda(r,t)$, where the index $\lambda$ assumes values $ss$, $st$, or $ts$.  The corresponding potentials (in lattice units) can be determined [\textit{cf.\/}\ Eq.\ (\ref{eq:W_at_large_t})] by fitting logarithms of the measured loops by linear functions in $t$:
\begin{equation}\label{eq:linear_fit}
-\ln W_\lambda(r,t)=C_\lambda+\widehat{V}_\lambda(r)\cdot t,\qquad\lambda\in\{ss, st, ts\},
\end{equation}
in an appropriate interval of $t$ values, from $t_\mathrm{min}$ to $t_\mathrm{max}$. Usually, to get a reasonable fit, one has to seek a compromise between two conflicting requirements: on one hand, to have a sufficiently long $t$-interval, \textit{i.e.\/}\ to use as small $t_\mathrm{min}$ and as large $t_\mathrm{max}$ as possible, and, on the other hand, not to use too large $t$ values to avoid loops measured with unacceptably large error bars. In practice, however, with smeared links the problem turned out to be less severe: logarithms of Wilson loops were to a good approximation linear down to $t=1$ (see below), so we could choose for our fits $t_\mathrm{min}=1$. The upper limit $t_\mathrm{max}$ was chosen as a value, at which the quantity used to monitor the ground-state overlap (see Appendix \ref{sec:overlap}) dropped considerably or its error was of the size of its value.

  A linear fit (\ref{eq:linear_fit}) in a fixed $(t_\mathrm{min},t_\mathrm{max})$-interval using the standard least-square method provides an expected value of $\widehat{V}_\lambda(r)$ together with an estimate of its statistical error, $\delta\widehat{V}_\lambda(r)$. This is almost certain to be an underestimate of the true error of the determination of 
the potential: there exists also a systematic error due to the unavoidably subjective nature of the choice of the fitting interval. To make at least a rough estimate of the order of magnitude of this systematic error, we used a kind of ``jackknife-like'' procedure: From the interval $(1,t_\mathrm{max})$ we omit in turn the 1-st, 2-nd, etc., $j$-th point and make a linear fit (\ref{eq:linear_fit}) to data points without the omitted point. In this way we obtain an ensemble of fit values for $\widehat{V}_\lambda(r)$, and estimate the systematic error of the procedure as the square root of the variance of this ensemble. It turns out that at small $r$ values the systematic error is of the order of the statistical one, but at large $r$ values it is often an order of magnitude larger.

	Finally, if we want to compare potentials at different values of the coupling $\beta$, we need to fix the physical scale. Usually, lattice spacings $a_s$ are expressed in units of the Sommer scale $r_0$ defined via the relation \cite{Sommer:1993ce}:
\begin{equation}\label{eq:sommer}
\left. \mathbf{r}^2\frac{dV}{d\mathbf{r}}\right|_{\mathbf{r}=r_0}=
\left. r^2\frac{d\widehat{V}}{dr}\right|_{r=r_0/a_s}=1.65.
\end{equation} 
In lattice QCD, one uses for the Sommer scale a phenomenological value of $r_0=0.5$ fm, in the case of \gtwo we leave this scale unfixed. In fact, for illustrative purposes it suffices to use approximate proportionality of  $a_s(\beta)/r_0$ to $\sqrt{\widehat\sigma(\beta)}$, where $\widehat\sigma(\beta)$ is the string tension measured in lattice units (see Figs.~\ref{fig:fund_phys} and \ref{fig:adj_phys} in Sec.~\ref{subsec:higher}). A small correction due to the Coulomb coupling constant does not change the picture qualitatively.  
	 
\subsection{Potential for fundamental representation}\label{subsec:fundamental}

\begin{table}[!t]
\caption{Parameters of the static potential between fundamental-representation charges, Eq.~(\ref{eq:potential_fit}),  symmetric lattice ($14^4$). The physical scale $a_s/r_0$ was set using Eq.~(\ref{eq:sommer}).}\label{tab:fund_sym} 
\begin{ruledtabular}
\begin{tabular}{c c c c c c}
&\multicolumn{3}{c}{Our results}&\multicolumn{2}{c}{Ref.~\cite{Greensite:2006sm}}\\\cline{2-4}\cline{5-6}
 $ \beta $ & $ \alpha $   &  $\widehat{\sigma}$ & $a_s/r_0$  
 & $\alpha$ &  $\widehat{\sigma}$ \\\hline
 9.5 &    0.291(8)  &  0.232(3) & 0.414(3) & 0.28(5)  & 0.24(2)  \\
 9.6 &    0.294(4)  &  0.148(1) & 0.331(2) & 0.313(2) & 0.14(1) \\
 9.7 &    0.291(4)  &  0.112(2) & 0.287(2) & 0.318(8) & 0.102(3)   
\end{tabular}
\end{ruledtabular}
\end{table} 
\begin{table*}[!t]
\caption{Parameters of the static potential between fundamental-representation charges, Eq.~(\ref{eq:potential_fit}), asymmetric lattice ($14^4\times 28$). The physical scale $a_s/r_0$ was set using Eq.~(\ref{eq:sommer}) from $\widehat{V}_{st}$.}\label{tab:fund_asym} 
\begin{ruledtabular}
\begin{tabular}{c c c c c c c c}
&\multicolumn{3}{c}{$\widehat{V}_{st}$}&\multicolumn{2}{c}{$\widehat{V}_{ss}$}&\multicolumn{2}{c}{$\widehat{V}_{ts}$}\\\cline{2-4}\cline{5-6}\cline{7-8}
 $ \beta $ & $ \alpha_{st} $   &  $\widehat{\sigma}_{st}$ & $a_s/r_0$
 & $\alpha_{ss}$ &  $\widehat{\sigma}_{ss}$ & $\alpha_{ts}$ &  $\widehat{\sigma}_{ts}$ \\\hline
 9.5 &  0.118(5)   &   0.1997(19) & 0.532(12)  &  0.249(31)  &  0.399(13)   & 0.194(5) &  0.2482(22) \\
 9.6 &  0.124(4)   &   0.1303(18) & 0.427(15)  &  0.264(34)  &  0.257(15)   & 0.231(7) &  0.1784(30) \\
 9.7 &  0.122(5)   &   0.0999(21) & 0.372(18)  &  0.264(36)  &  0.196(15)   & 0.244(8) &  0.1452(33) 
\end{tabular}
\end{ruledtabular}
\end{table*} 
\begin{figure}[!t]
\begin{center}
\includegraphics[width=8truecm]{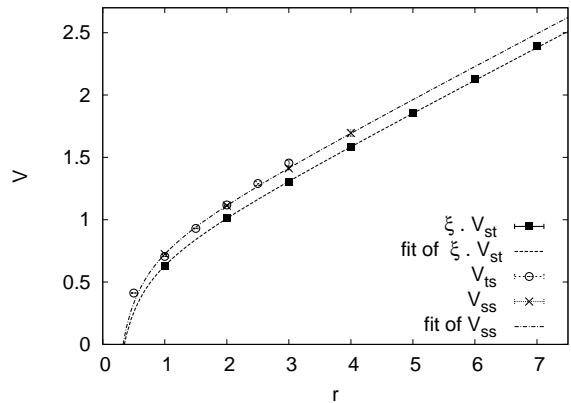}
\end{center}
\caption{Potentials $\widehat{V}_{st} $, $\widehat{V}_{ss} $, and $\widehat{V}_{ts} $. Distances $r$ are measured in units of $a_s$. ($14^3 \times 28$-lattice, $ \beta = 9.6 $, $\xi = 1.97$.)}\label{fig:pot_96plusfit}
\end{figure}
\begin{figure}[!t]
\begin{center}
\includegraphics[width=8truecm]{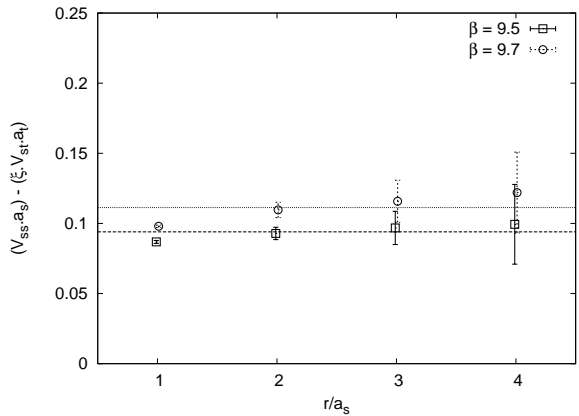}
\end{center}
\caption{Differences between potentials $\widehat{V}_{ss}$ and $\widehat{V}_{st}$ (multiplied by $\xi$). ($14^3 \times 28$-lattice: $\beta = 9.5 $, $\xi = 2.00$; $\beta = 9.7 $, $\xi = 1.96$.)}\label{fig:distpot}
\end{figure}

	In this subsection we summarize selected results for potentials between static quarks and antiquarks from the fundamental representation, and the main purpose is to calibrate our procedure and compare with results of an earlier investigation \cite{Greensite:2006sm}. Results for higher representations are deferred until Sec.~\ref{subsec:higher}.
	
	Our simulations were performed at three values of the coupling $\beta$ on the weak coupling side of the
crossover region observed in Ref.\ \cite{Pepe:2006er}. Parameters of the production runs are summarized in Table~\ref{tab:runs}.

	Fig.~\ref{fig:potbf14_96_st} clearly illustrates the fact that has been mentioned in Sec.~\ref{subsec:fits}: logarithms of Wilson loops computed from smeared links are almost perfect linear functions of~$t$. The example shows $W_{st}(r,t)$ for a series of $t$ values at fixed $r$. The corresponding value of the potential was then obtained by a fit (\ref{eq:linear_fit}). The resulting potentials were then parametrized by the usual Coulomb plus linear form:
\begin{equation}\label{eq:potential_fit}
\widehat{V}_\lambda(r)=\widehat{c}_\lambda-\frac{\alpha_\lambda}{r}+\widehat\sigma_\lambda\; r,\qquad\lambda\in\{ss, st, ts\}.
\end{equation}
(A hat over a number again indicates that it is dimensionless lattice quantity, no hat is needed on $\alpha$.)

	To compare with results in the existing literature, we computed $\alpha$ and $\widehat\sigma$ in runs on a symmetric, $14^4$ lattice. The result is summarized in Table~\ref{tab:fund_sym}. The values of the string tension are in reasonable agreement with those of Ref.\ \cite{Greensite:2006sm}, obtained on $12^4$ lattice. The slight discrepancy between $\alpha$ values can be attributed to different smearing procedures adopted. (Ref.\ \cite{Greensite:2006sm} used a variant of the procedure of \cite{Langfeld:2003ev}.) The purpose of smearing is to reduce ultraviolet noise in lattice configurations that mostly influences the self-energy and Coulombic parts of the static potential, and efficiency of various procedures differs in achieving this goal. At $\beta=9.7$ also string tensions seem not to be compatible within errors. However, both numbers are quoted with statistical errors only, and may in fact agree if finite-size effects and/or systematic errors are taken into account. To support this claim, we determined the string tension by our procedure also on $12^4$ lattice, and  obtained a value of 0.109(2). Finally, using $t_{\mathrm{min}}=2$ instead of 1 in potential fits (\textit{cf.}\ the discussion in Sec.~\ref{subsec:fits}), we went further down to $\widehat\sigma=0.105(2)$. It should be clear that though statistical errors on string tensions are as low as 2--3\%, the actual error may be of the order of 10\%.  

	Next we turn to asymmetric lattices and determine parameters of the three potentials (\ref{eq:potss}), (\ref{eq:potst}), and	(\ref{eq:potts}). Those are summarized in Table~\ref{tab:fund_asym}.\footnote{We have some data on static potentials for the fundamental representation also from simulations on a $16^3\times 32$ lattice, albeit with lower statistics. The estimated string tensions agree with those quoted above, no strong finite-size effect was observed in this case. \textit{E.g.}, our preliminary values from the $16^3\times 32$ lattice are $\widehat\sigma_{st}= 0.199(2), 0.129(2), 0.097(2)$ at $\beta=9.5, 9.6, 9.7$, respectively.} When looking at these numbers, one should keep in mind that the errors quoted are just statistical. We did not attempt to estimate systematic errors, which were discussed in Sec.~\ref{subsec:fits}. The most reliable determination of parameters of the static potential comes from Wilson loops $W_{st}$ with the $t$-direction of the loop along the long lattice direction (in which the lattice is more finely discretized). 
	
	One can now verify the correctness of the chosen values for the bare anisotropy $\xi_0$ (\textit{cf.} Table~\ref{tab:xi_0}). On the basis of Eq.~(\ref{eq:pot_rel1}) one expects $\widehat\sigma_{ss}\approx 2\widehat\sigma_{st}$ and $\alpha_{ss}\approx 2\alpha_{st}$, which is satisfied to a surprising accuracy. From the numerical values of the string tensions one obtains the renormalized anisotropies $\xi= 2.00(7), 1.97(12),$ and $1.96(16)$ for $\beta=9.5, 9.6$, and $9.7$, respectively. Similar relations should hold between \textit{ss} and \textit{ts} parameters [see Eq.~(\ref{eq:pot_rel2})]. However, they are fulfilled with worse accuracy, because of systematic errors mentioned above. The available $t$ intervals for determination of these potentials are short, since the overlap with the ground state drops fast with $t$ at larger values of $r$ (see Fig.~\ref{fig:ov96sm} in Appendix~\ref{sec:overlap}).
	
	The fulfillment of relations~(\ref{eq:pot_rel1}) and (\ref{eq:pot_rel2}) is visually displayed in Fig.~\ref{fig:pot_96plusfit} for $\beta=9.6$. The difference between $\widehat{V}_{ss}$ and $\xi\;\widehat{V}_{st}$ is expected to be a constant independent of $r$ (difference between self energies), our data support the expectation, see Fig.~\ref{fig:distpot}.
	
	Finally, we checked that the exact value of the bare anisotropy $\xi_0$ is not of vital importance on asymmetric lattices. There were no large differences between the obtained potentials and their string tensions when we used $\xi_0=1.61$ instead of 1.59 at $\beta=9.5$, or $\xi_0=1.62$ instead of 1.64 at $\beta=9.7$.
	
\subsection{Higher representations}\label{subsec:higher}

\begin{figure*}[!tb]
\begin{tabular}{cc}
\includegraphics[height=8truecm,angle=-90]{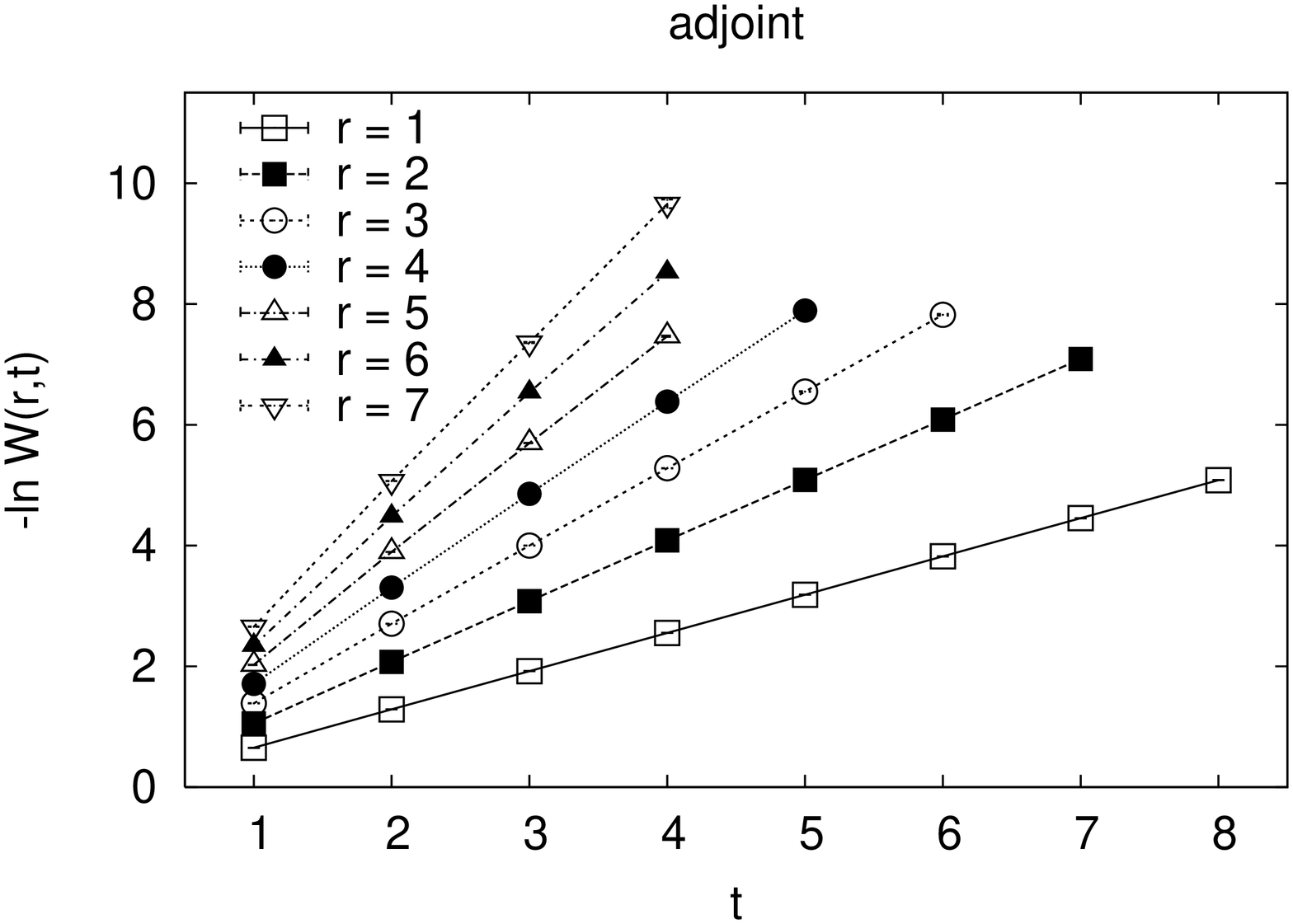} &
\includegraphics[height=8truecm,angle=-90]{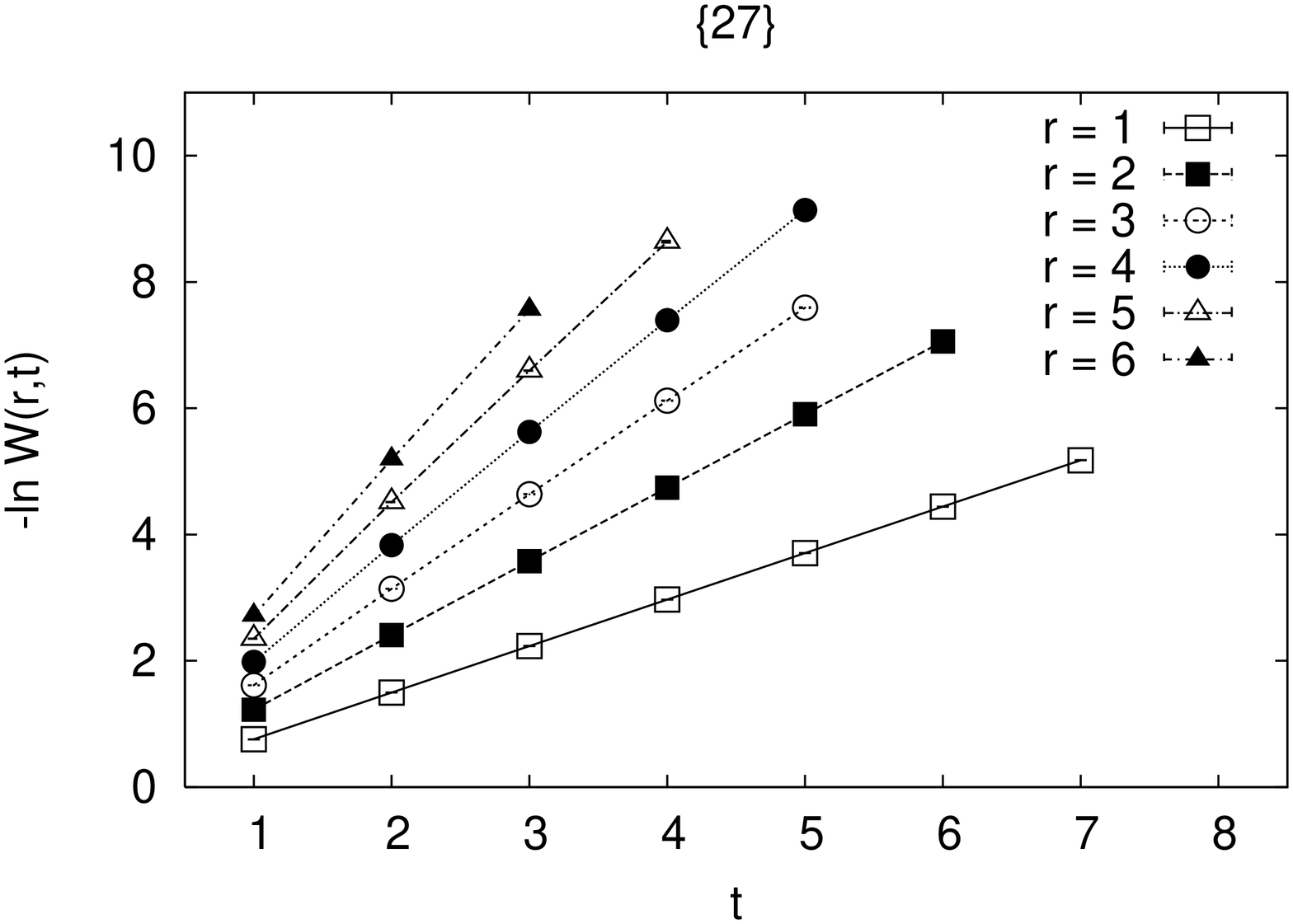}\\
\includegraphics[height=8truecm,angle=-90]{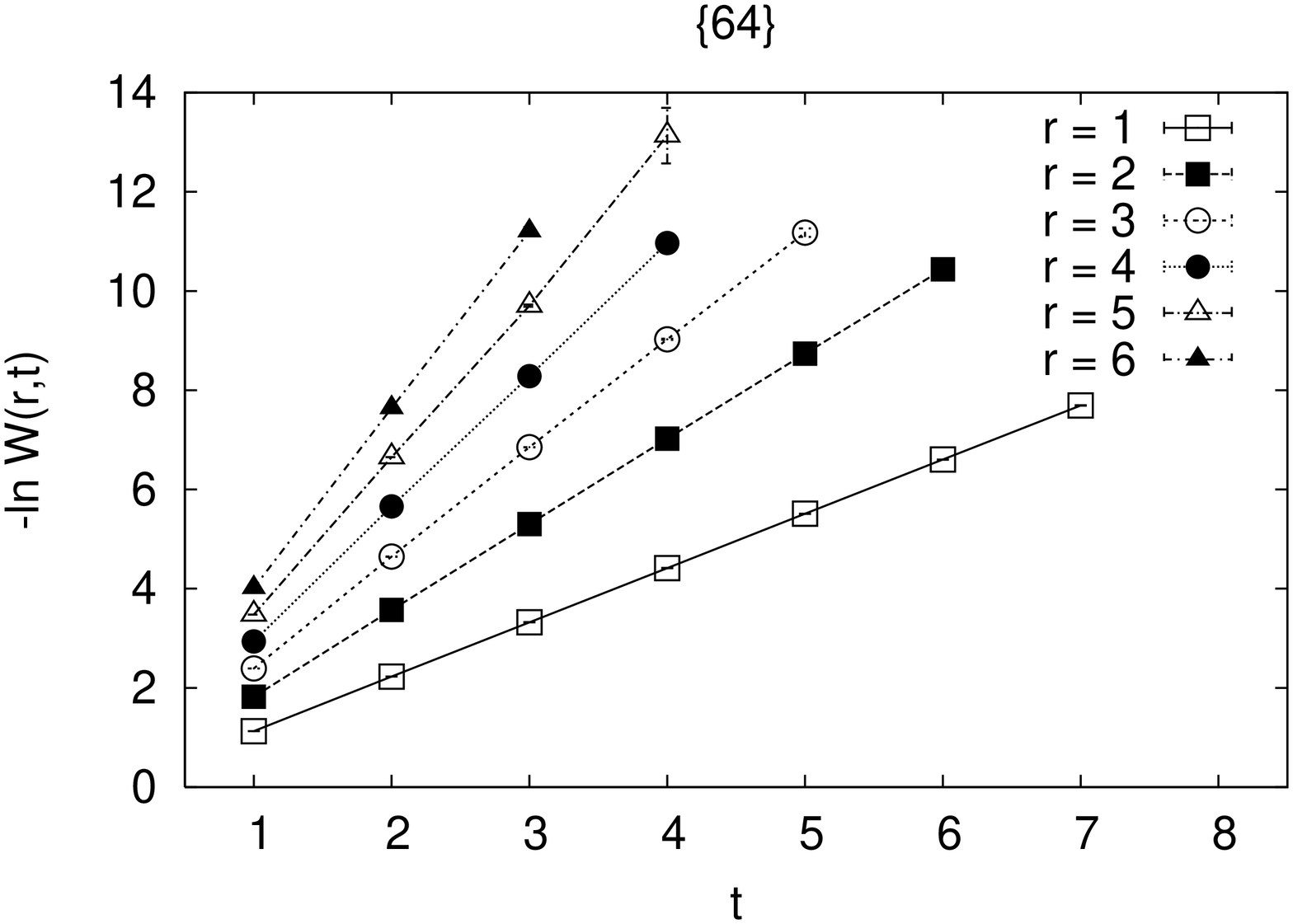} &
\includegraphics[height=8truecm,angle=-90]{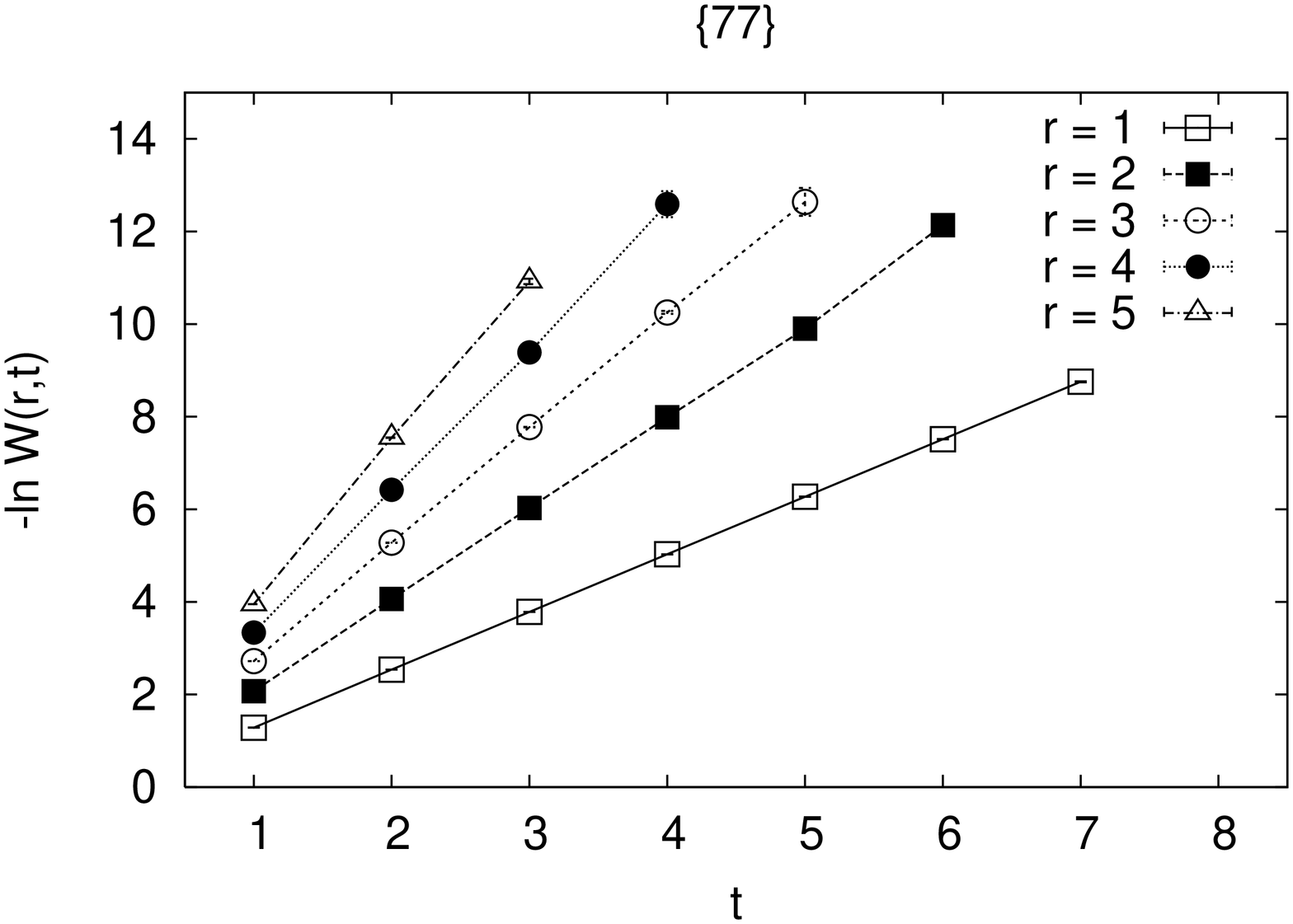}
\end{tabular}
\caption{Logarithm of the expectation value of the Wilson loop $W_{st}(r,t)$ for fixed $r$-values as a function of $t$, in the adjoint, $\{27\}$-, $\{64\}$-, and $\{77\}$-representations. ($14^3 \times 28$ lattice, $\beta = 9.6$. Lines are drawn to guide the eye.)}\label{fig:ln_higher}
\end{figure*}

\begin{figure}
\centering
\includegraphics[width=8truecm]{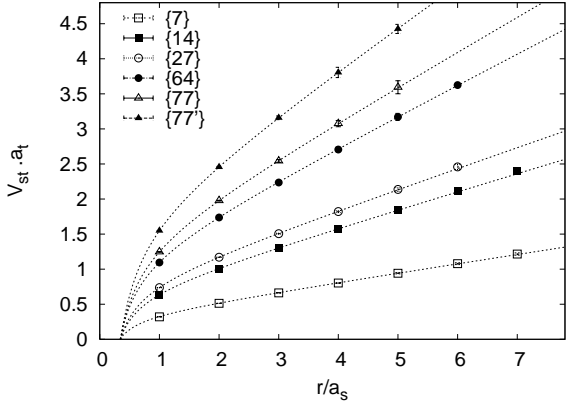}
\caption{The potentials $\widehat{V}_{st}$ for different representations vs.\ the dimensionless $r$. ($14^3 \times 28$ lattice, $ \beta = 9.6$.)}\label{fig:potens96}
\end{figure}

\begin{table*}[!tb]
\caption{String tensions $\widehat{\sigma}_{st}$ for different representations. ($14^3 \times 28$ lattice. Because of large errorbars, it was not possible to determine reliably string tensions for the $\{77\}$ and $\{77'\}$ representations at $\beta=9.5$.)} 
\begin{ruledtabular}\label{tab:tensions_higher}
\begin{tabular}[c]{ c  c c  c c  c c }      
 & \multicolumn{2}{c}{fundamental} & \multicolumn{2}{c}{adjoint} &  \multicolumn{2}{c}{$\{27\}$}  \\
\cline{2-3}\cline{4-5}\cline{6-7}
 $\beta$ & $\widehat\sigma$ &  $\widehat\sigma/\widehat\sigma_F$ & $\widehat\sigma$ &  $\widehat\sigma/\widehat\sigma_F$ & $\widehat\sigma$ &  $\widehat\sigma/\widehat\sigma_F$ \\\hline
 9.5         &    0.1997(19) & 1 & 0.376(6) & 1.883(35) &  0.429(8) & 2.148(45) \\
 9.6         &    0.1303(18)  & 1 &  0.253(4)  & 1.942(40) &  0.292(6) & 2.241(55) \\
 9.7         &    0.0999(21)  & 1 &  0.196(4)  & 1.962(57)  &  0.228(5) & 2.282(69)  \\
 \hline
\multicolumn{2}{l}{Casimir ratio:}& 1 & & 2 & & 2.333   \\\hline\hline
 & \multicolumn{2}{c}{$\{64\}$}  &  \multicolumn{2}{c}{$\{77\}$} &  \multicolumn{2}{c}{$\{77 ' \}$}  \\
\cline{2-3}\cline{4-5}\cline{6-7}
 $\beta$ & $\widehat\sigma$ &  $\widehat\sigma/\widehat\sigma_F$ & $\widehat\sigma$ &  $\widehat\sigma/\widehat\sigma_F$ &  $\widehat\sigma$ &  $\widehat\sigma/\widehat\sigma_F$\\\hline
 9.5         & 0.612(25) & 3.065(129) & --- & ---   & --- &   --- \\
 9.6         &  0.436(8) & 3.346(77) & 0.488(20) & 3.745(162) &  0.596(22) & 4.574(180) \\
 9.7         &  0.347(11)  & 3.473(132) &  0.396(16)  & 3.964(181)  &  0.492(16) & 4.925(191) \\
 \hline
\multicolumn{2}{l}{Casimir ratio:}& 3.5 & & 4 & & 5            
\end{tabular}
\end{ruledtabular}
\end{table*}
\begin{figure}[!tb]
\begin{center}
\includegraphics[width=8truecm]{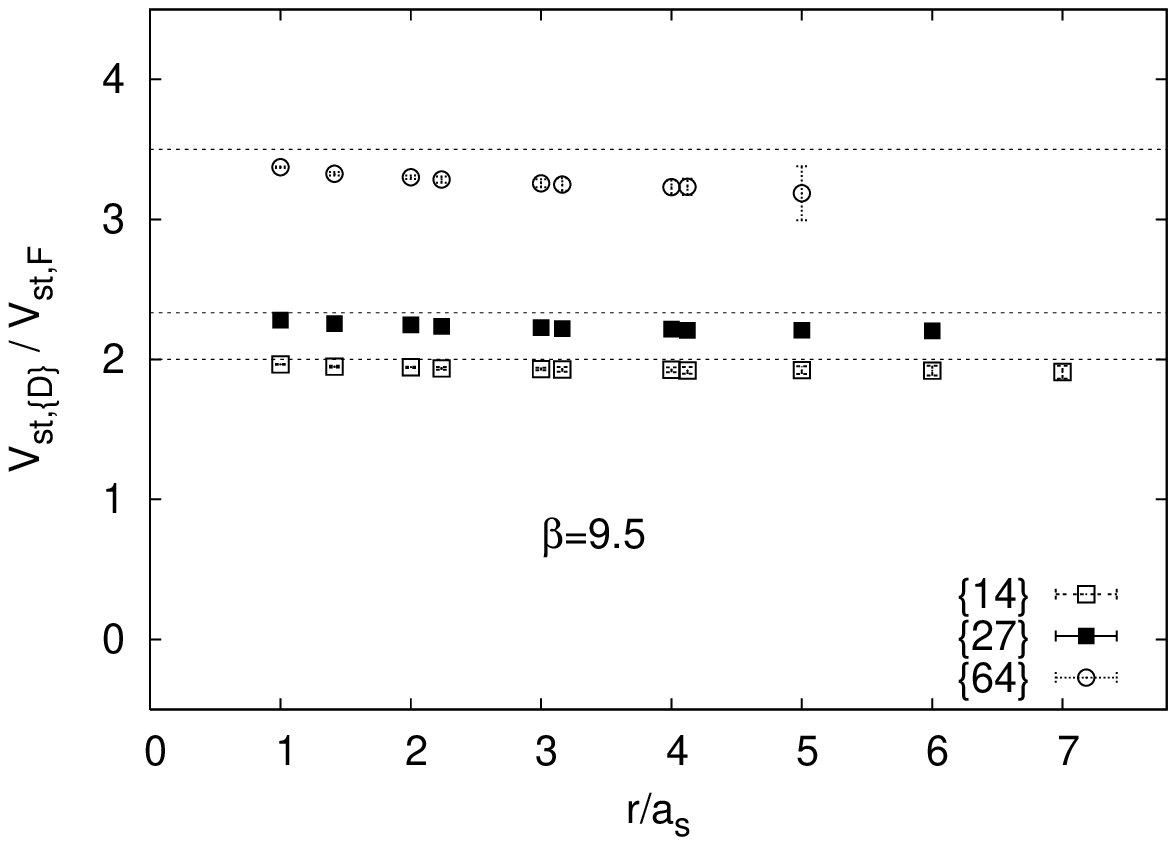} \\
\includegraphics[width=8truecm]{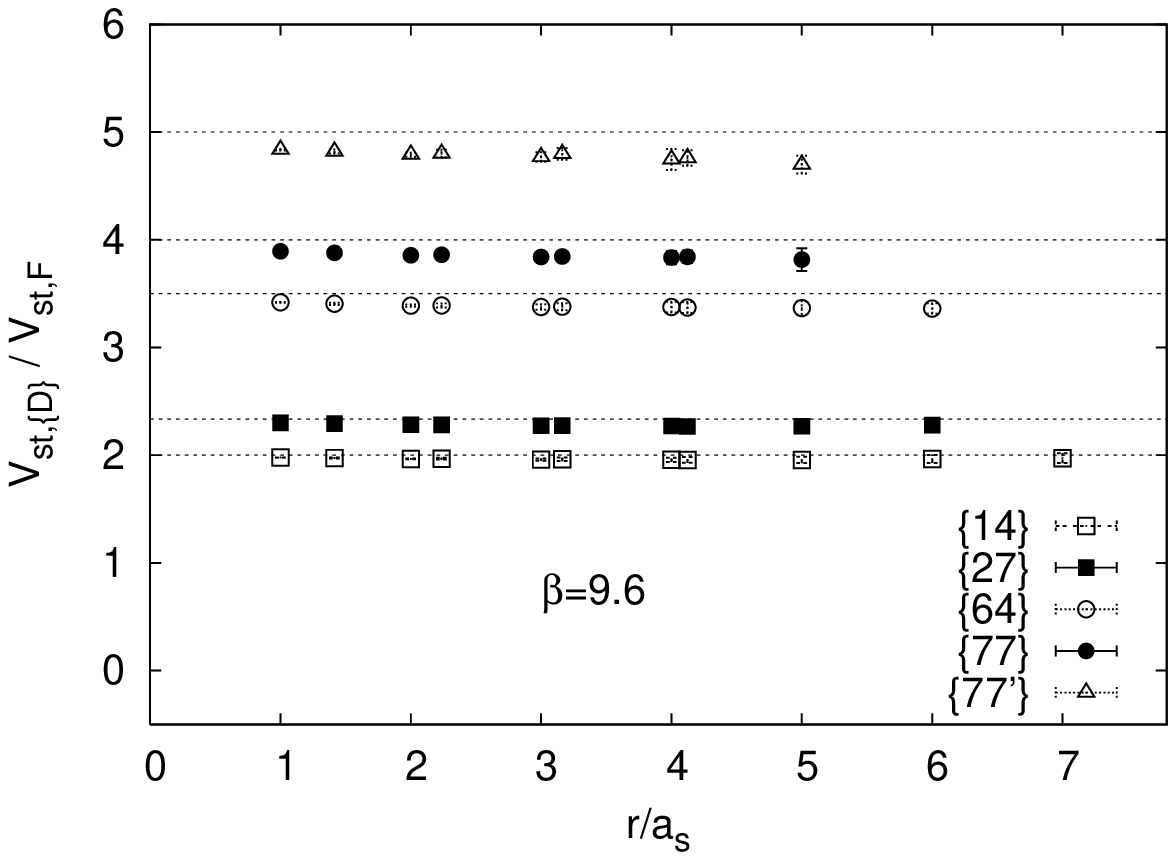} \\
\includegraphics[width=8truecm]{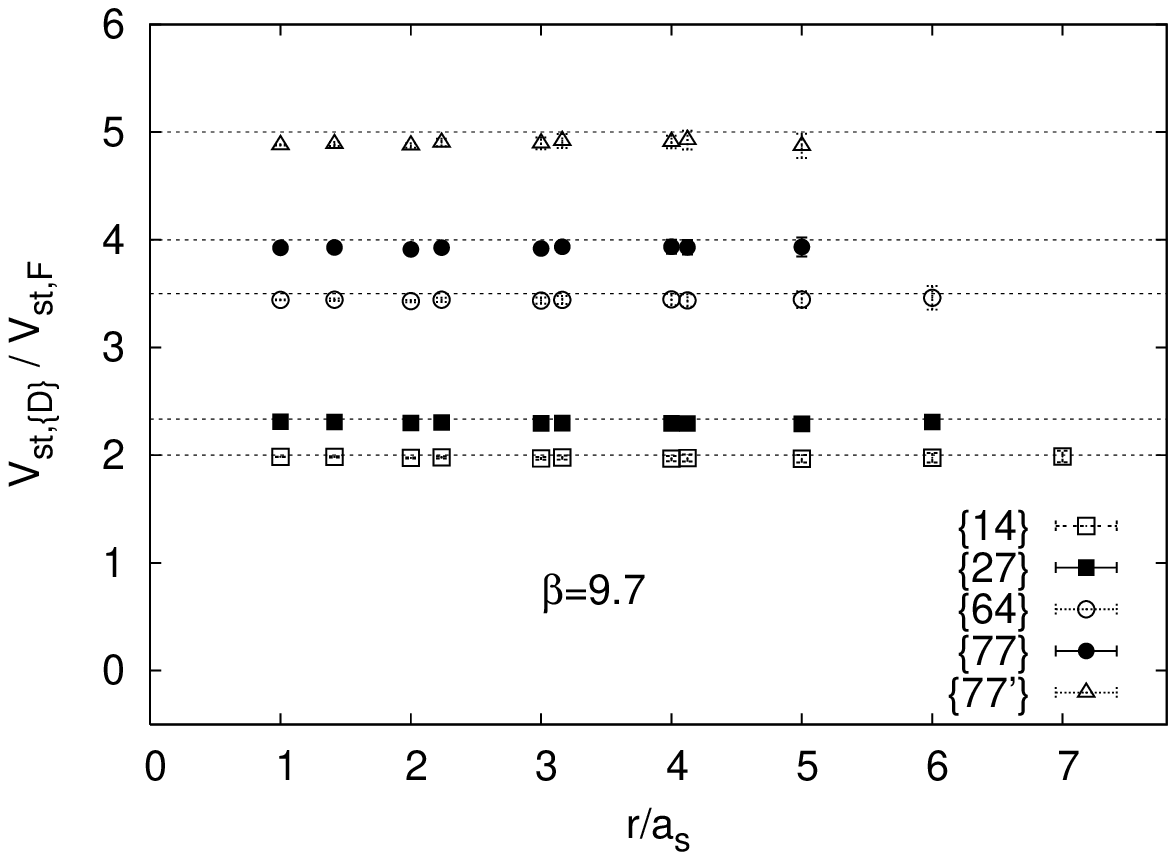} \\
\end{center}
\caption{Ratios $\widehat{V}_{st,\{D\}} / \widehat{V}_{st,F}$ for different representations $\{D\}$ are shown as functions of the dimensionless $r$. Horizontal lines show Casimir-scaling predictions. ($14^3 \times 28$ lattice. A few points from off-axis Wilson loops are also included, those were not used in string-tension determination.)}\label{fig:ratios}
\end{figure}
\begin{figure}[!tb]
\begin{center}
\includegraphics[width=8truecm]{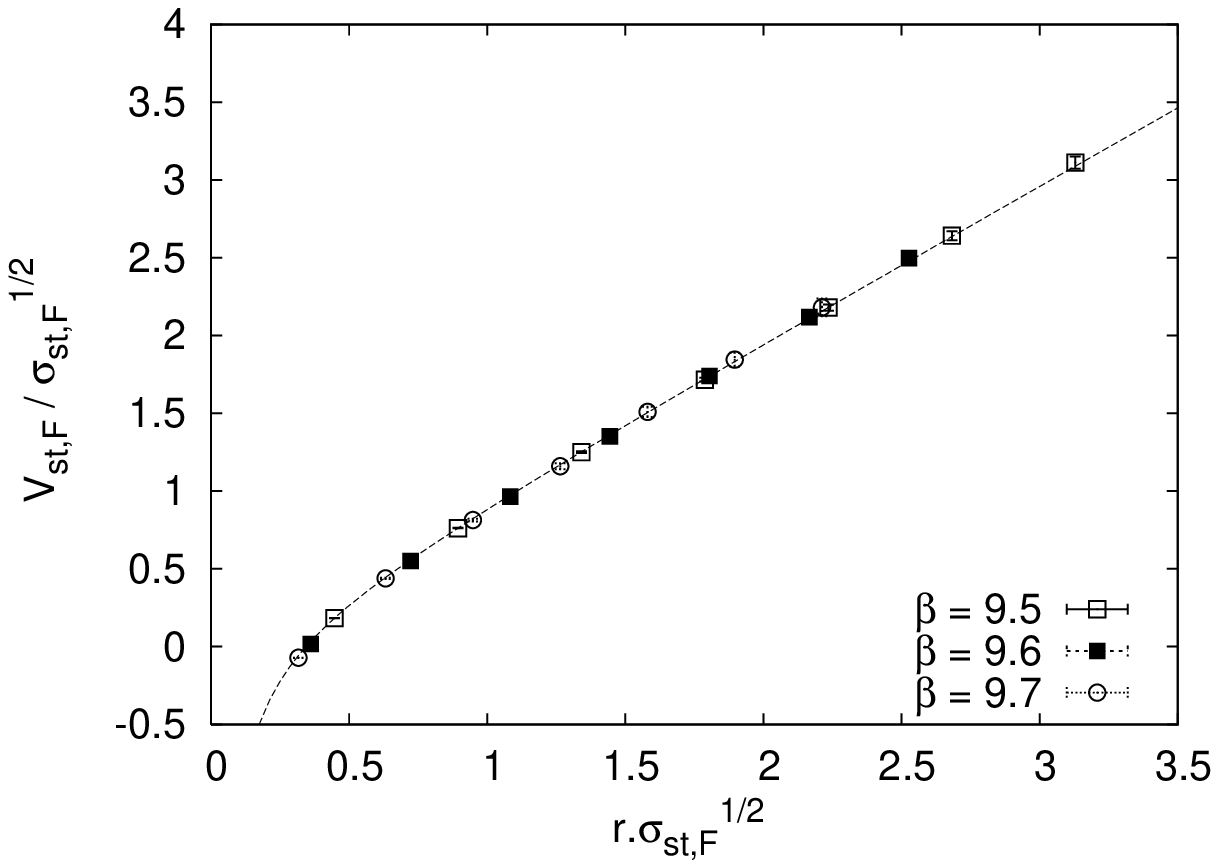}
\end{center}
\caption{The fundamental-representation potential in physical units, all $\beta$ values, from $14^3\times 28$ lattice. (The constant $\widehat{c}$, \textit{cf.}\ Eq.~(\ref{eq:potential_fit}), is subtracted. The line is drawn to guide the eye.)}\label{fig:fund_phys}
\end{figure}
\begin{figure}[!tb]
\begin{center}
\includegraphics[width=8truecm]{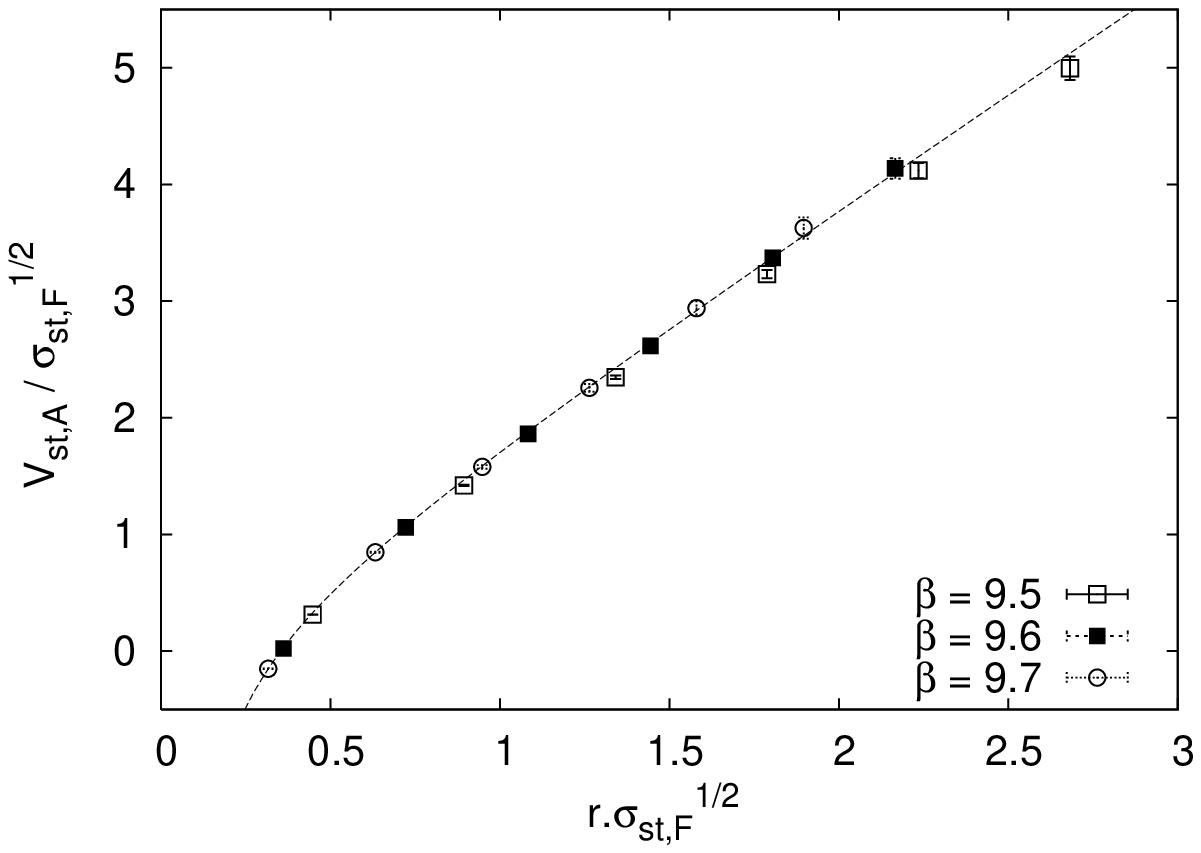}
\end{center}
\caption{The adjoint-representation potential in physical units, all $\beta$ values, from $14^3\times 28$ lattice. (The constant $\widehat{c}$, \textit{cf.}\ Eq.~(\ref{eq:potential_fit}), is subtracted. The line is drawn to guide the eye.)}\label{fig:adj_phys}
\end{figure}

	After subjecting our procedure for determination of potentials to a number of tests in the case of the fundamental representation, we are ready to proceed to the main subject of the present paper: potentials between static color sources from higher representations of the \gtwo group. For this purpose we will concentrate on potentials $\widehat{V}_{st}$ which can be determined most accurately. The procedure is now in principle straightforward. A thermalized configuration on an $L^3\times(2L)$ lattice is a set of link matrices in the fundamental representation. We smeared these links over ``short'' directions using the procedure described in Sec.~\ref{subsec:smearing}. Then, we compute a fundamental Wilson loop as a product of smeared link matrices along a closed loop (one side in a ``short'' direction, the other in the ``long'' one). Before taking a trace, we compute the matrix representing this loop in the adjoint representation using Eq.\ (\ref{eq:fun_to_adj}). Then we trace both the fundamental and adjoint Wilson loop matrix, and from these traces determine also those of the loop matrices in higher representations via the set of formulas (\ref{eq:trace_formulas}). Finally, to obtain the expectation values of traces of Wilson loops of size ($r\times t$), we average over loop positions on a lattice, and over an ensemble of thermalized configurations in the usual way.
	
	Illustrative results for Wilson loops in higher representations are displayed in Fig.~\ref{fig:ln_higher} for $\beta=9.6$. The results look very much the same as in the case of the fundamental representation (Fig.~\ref{fig:potbf14_96_st}). Values of the potential at various separations $r$ are obtained by fits of the form (\ref{eq:linear_fit}). The results for $\beta=9.6$ are shown in Fig.~\ref{fig:potens96}. It is clear that all higher potentials show a linear rise similar to the fundamental representation in the same range of distances.\footnote{Though the string breaks in all representations of G${}_2$, we had no ambition to observe it in our simulations. It is well known from earlier studies (see \textit{e.g.}\ \cite{Kratochvila:2003zj}) that it is of utmost difficulty to see string breaking on a lattice. Such a task is currently far beyond our possibilities, and is also not relevant to the main message of the paper.}

	The obtained potentials were fitted again by the Coulomb plus linear form (\ref{eq:potential_fit}). The resulting values of string tensions are summarized in Table~\ref{tab:tensions_higher}. The measured values lie very close to predictions based on Casimir scaling, and the agreement with Casimir scaling tends to improve by increasing $\beta$, \textit{i.e.} the deeper we enter the weak-coupling region above the crossover. In Fig.~\ref{fig:ratios} this fact is exemplified on ratios of different representation potentials to the fundamental one at fixed $r$. At all distances, the ratios are close to the Casimir-scaling prediction, even though we made no attempt to get rid of the self-energy contributions.

	In this context, the question of systematic errors again pops up: Could they not invalidate the above agreement with Casimir scaling? As discussed in detail in Sec.\ \ref{subsec:fits}, the main source of the systematic error is the choice of the $t$-interval for linear fits of the form (\ref{eq:linear_fit}). To produce results in Table \ref{tab:tensions_higher}, $t_{\mathrm{min}}=1$ was used. The same analysis with $t_{\mathrm{min}}=2$ for the three lowest representations leads to values given in Table \ref{tab:tensions_higher_2}. The Casimir-scaling result comes out to be quite robust against this modification: though string tensions changed in all representations by up to 10\%, they went in the same direction, and thus their ratios changed only slightly. In fact, at the largest available $\beta$, the agreement with Casimir-scaling prediction even improved.  

\begin{table*}[!tb]
\caption{Same as Table \protect\ref{tab:tensions_higher}, from potential fits (\ref{eq:linear_fit}) with $t_{\mathrm{min}}=2$.} 
\begin{ruledtabular}\label{tab:tensions_higher_2}
\begin{tabular}[c]{ c  c c  c c  c c }      
 & \multicolumn{2}{c}{fundamental} & \multicolumn{2}{c}{adjoint} &  \multicolumn{2}{c}{$\{27\}$}  \\
\cline{2-3}\cline{4-5}\cline{6-7}
 $\beta$ & $\widehat\sigma$ &  $\widehat\sigma/\widehat\sigma_F$ & $\widehat\sigma$ &  $\widehat\sigma/\widehat\sigma_F$ & $\widehat\sigma$ &  $\widehat\sigma/\widehat\sigma_F$ \\\hline
 9.5         &    0.192(1)  & 1 &  0.358(6)  & 1.866(34)  &  0.406(8) & 2.120(43) \\
 9.6         &    0.121(1)  & 1 &  0.237(3)  & 1.960(30)  &  0.273(4) & 2.255(41) \\
 9.7         &    0.090(1)  & 1 &  0.181(3)  & 2.015(40)  &  0.209(4) & 2.328(54)  \\
 \hline
\multicolumn{2}{l}{Casimir ratio:}& 1 & & 2 & & 2.333          
\end{tabular}
\end{ruledtabular}
\end{table*}

	Finally, we display the static potential for color charges in the fundamental (Fig.~\ref{fig:fund_phys}) and adjoint representation (Fig.~\ref{fig:adj_phys}) together for all three $\beta$ values, expressed in physical units defined by the fundamental string tension. All potentials nicely follow a universal curve.
	
\section{Conclusions}\label{sec:conclusions}

	Alice, in Lewis Carroll's \textit{Through the Looking-Glass, and What Alice Found There\/}, enters a garden, where flowers speak to her and mistake her for a flower. Does the gauge theory with the exceptional group \gtwo belong to the same species of flowers with ``ordinary'' confining SU($N$) gauge theories, or is it different and only mistaken for a flower? Many results of recent investigations suggest the answer that all non-Abelian gauge theories, including G${}_2$, share many properties in the infrared. Thermodynamic properties are similar~\cite{Pepe:2006er}, and Ref.\ \cite{Maas:2007af} provided further numerical evidence: a qualitative agreement among Landau-gauge ghost and gluon propagators and the Faddeev--Popov operator eigenspectrum in SU(2), SU(3) and \gtwo gauge theories, though only in two and three Euclidean dimensions.

	From the point of view of the center-vortex confinement mechanism, theories also behave similarly. When center vortices are present and percolate, in theories with non-trivial center, they cause the expected $N$-ality dependence of asymptotic string tensions for static potentials for color charges from different representations of the gauge group. In a theory with trivial center, the asymptotic string tension for all representations is zero. But even if a centerless theory possesses no non-trivial center-vortex-like objects, its vacuum can contain domain structures. Ref.\ \cite{Greensite:2006sm} suggested a universal and simple model in which such vacuum domains -- assuming color magnetic fields fluctuating almost independently in each domain, only fulfilling a constraint that the total flux corresponds to an element of the center -- cause Casimir scaling of static potentials in the intermediate distance interval, until color screening sets on. Casimir scaling was observed in lattice simulations of SU(2) and SU(3) gauge theories long ago, the model prediction called for verification also in the \gtwo lattice gauge theory.
	
	Our results convincingly demonstrate (approximate) Casimir scaling for static potentials between color charges from the six lowest representations of G${}_2$. The agreement between measured values of intermediate string tensions with predictions based on values of quadratic Casimirs is quite striking, they differ by at most 10--15\%, and this can hardly be just a numerical coincidence. The results of course cannot \textit{prove\/} that the model is right, but combined with the solid evidence for Casimir scaling in SU(2) and SU(3) provide support for its main ingredients (common for all groups) -- magnetically disordered vacuum with a domain structure. How to identify such domains is another question; in theories with non-trivial center one can apply the method of center projection in the maximal center gauge~\cite{DelDebbio:1998uu}, thus fitting extended (``thick'') vacuum structures by thin center vortices. The identification of vacuum-type domains with trivial total flux remains problematic.
	
	The observed Casimir scaling, coming out so naturally within models with magnetic disorder and vacuum domain structure, may pose a challenge to approaches that invoke disparate ideas to explain the phenomenon of confinement.

\acknowledgments{%
This research was supported by the Slovak Grant Agency for Science, Project VEGA No.\ 2/6068/2006.

{\v{S}}.O.\ is grateful to Jeff Greensite, Kurt Langfeld, and Axel Maas for collaboration on topics related to the present work, and numerous discussions on \gtwo and Casimir scaling. He would also like to acknowledge correspondence with Gunnar Bali and Sedigheh (Mina) Deldar about their SU(3) results, and Dan Zwanziger for sending him the electronic version of Piccioni's PhD thesis.}  
 	 
\appendix

\section{\gtwo generators}\label{sec:generators}

The \gtwo algebra for our representation (based on \cite{Macfarlane:2002hr}) is defined by the following generators:
\begin{eqnarray}
t_{1} & = & \frac{1}{\sqrt{8}} \left( P_{12} + P_{21} - P_{56} - P_{65}   \right), \nonumber \\
t_{2} & = & \frac{i}{\sqrt{8}} \left( P_{21} - P_{12} -  P_{56} +  P_{65}   \right) , \nonumber \\
t_{3} & = & \frac{1}{\sqrt{8}} \left( P_{11} - P_{22} - P_{55} + P_{66}   \right), \nonumber \\
t_{4} & = & \frac{1}{\sqrt{8}} \left( P_{13} + P_{31} -  P_{57} -  P_{75}   \right) , \nonumber \\
t_{5} & = & \frac{i}{\sqrt{8}} \left( P_{31}  - P_{13} - P_{57} +  P_{75}   \right), \nonumber \\
t_{6} & = & \frac{1}{\sqrt{8}} \left( P_{23} + P_{32} - P_{67} - P_{76}   \right) , \nonumber \\
t_{7} & = & \frac{i}{\sqrt{8}} \left( P_{32} - P_{23} - P_{67} + P_{76}   \right), \nonumber \\
t_{8} & = &  \frac{1}{\sqrt{24}} \left( P_{11}+ P_{22} - 2 P_{33}  \right.\nonumber \\
& - & \left. P_{55} - P_{66} + 2 P_{77}   \right) ,  \nonumber \\
t_{9}  & = &  \frac{1}{\sqrt{24}} \left( P_{16} - P_{25} + \sqrt{2} P_{34} + \sqrt{2} P_{43} \right.\nonumber \\
& - & \left. \sqrt{2} P_{47} - P_{52} + P_{61} - \sqrt{2} P_{74} \right) , \\
t_{10} &  = &  \frac{i}{\sqrt{24}} \left( P_{25} - P_{16} + \sqrt{2} P_{34} - \sqrt{2} P_{43} \right.\nonumber \\
& - & \left. \sqrt{2} P_{47} - P_{52} + P_{61} + \sqrt{2} P_{74}  \right) , \nonumber \\
t_{11} &  = &  \frac{1}{\sqrt{24}} \left( \sqrt{2} P_{24} - P_{17} + P_{35} + \sqrt{2} P_{42} \right.\nonumber \\
& - & \left. \sqrt{2} P_{46} + P_{53} - \sqrt{2} P_{64} - P_{71}  \right) , \nonumber \\
t_{12} &  = &  \frac{i}{\sqrt{24}} \left( P_{35} - P_{17} - \sqrt{2} P_{24} + \sqrt{2} P_{42} \right.\nonumber \\
& + & \left. \sqrt{2} P_{46} - P_{53} - \sqrt{2} P_{64} + P_{71}  \right) , \nonumber \\
t_{13} &  = &  \frac{1}{\sqrt{24}} \left( \sqrt{2} P_{14} +  P_{27} -  P_{36} + \sqrt{2}  P_{41} \right.\nonumber \\
& - & \left. \sqrt{2} P_{45} - \sqrt{2} P_{54} - P_{63} + P_{72}  \right) , \nonumber \\
t_{14} &  = &  \frac{i}{\sqrt{24}} \left( \sqrt{2} P_{14} - P_{27} + P_{36} - \sqrt{2} P_{41} \right.\nonumber \\
& - & \left. \sqrt{2} P_{45} + \sqrt{2} P_{54} - P_{63} + P_{72}  \right) , \nonumber
\end{eqnarray}
where $ (P_{ij})_{ab} = \delta_{ia} \delta_{jb} $. The generators are normalized to satisfy:
\begin{equation}
\tr\left(t_a \cdot t_b \right)  = {\textstyle\frac{1}{2}} \; \delta_{ab}.
\end{equation}
%

\section{Overlap with the ground state}\label{sec:overlap}

\begin{figure}[t!]
\includegraphics[width=8truecm]{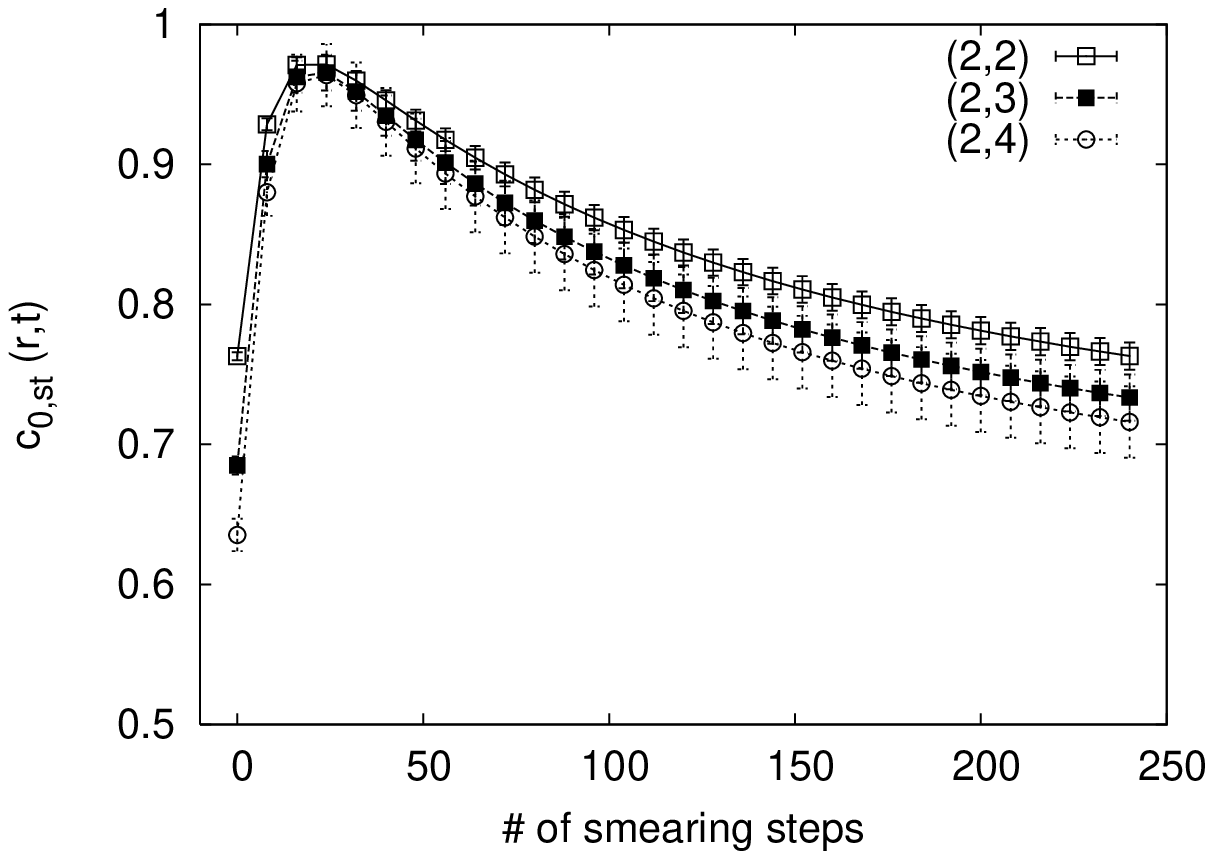} \\
\includegraphics[width=8truecm]{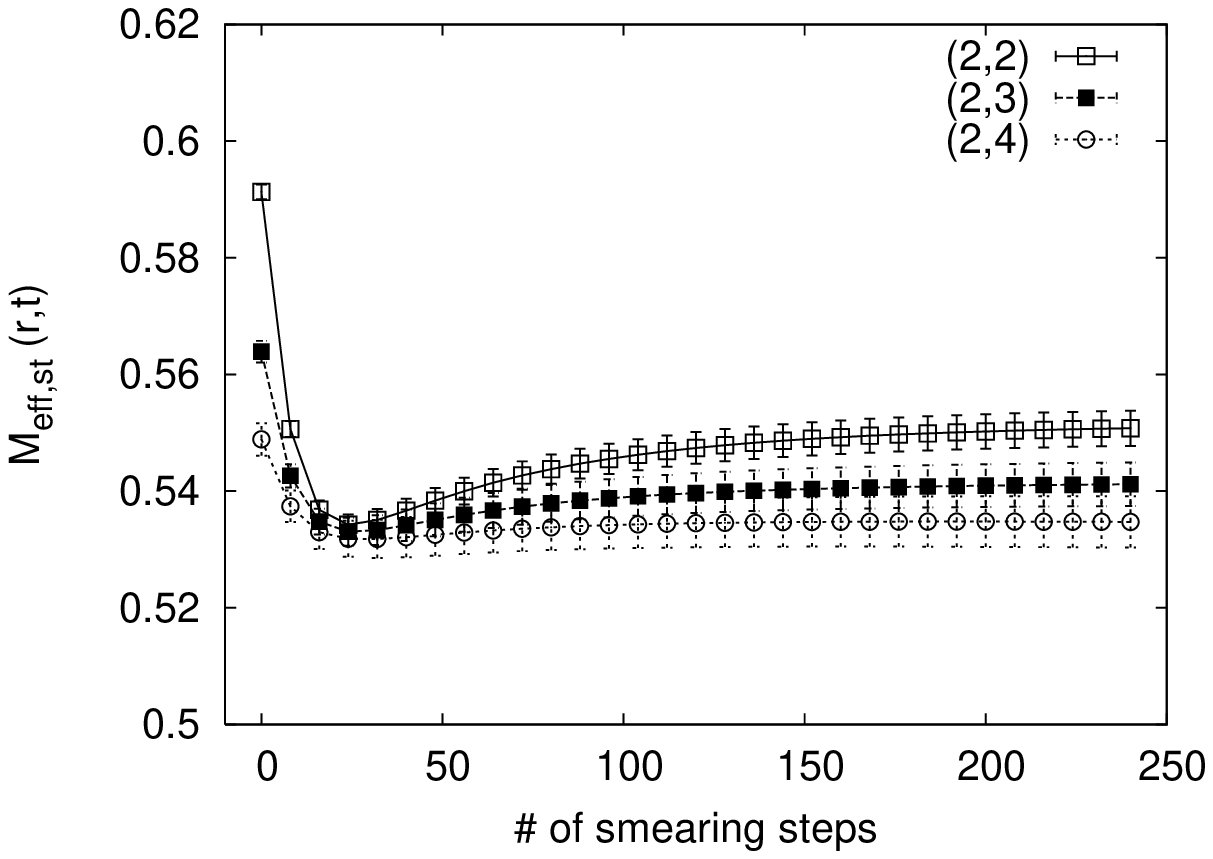}  
\caption{The quantities $c_{0,st}(r,t)$ and $M_{\mathrm{eff},st} (r, t)$ as functions of the number of smearing steps. ($12^3 \times 24$ lattice, $\beta = 9.5$, $\xi_0=1.8$.)}
\label{fig:klur01}
\end{figure}
\begin{figure}[!tbhp]
\includegraphics[width=8truecm]{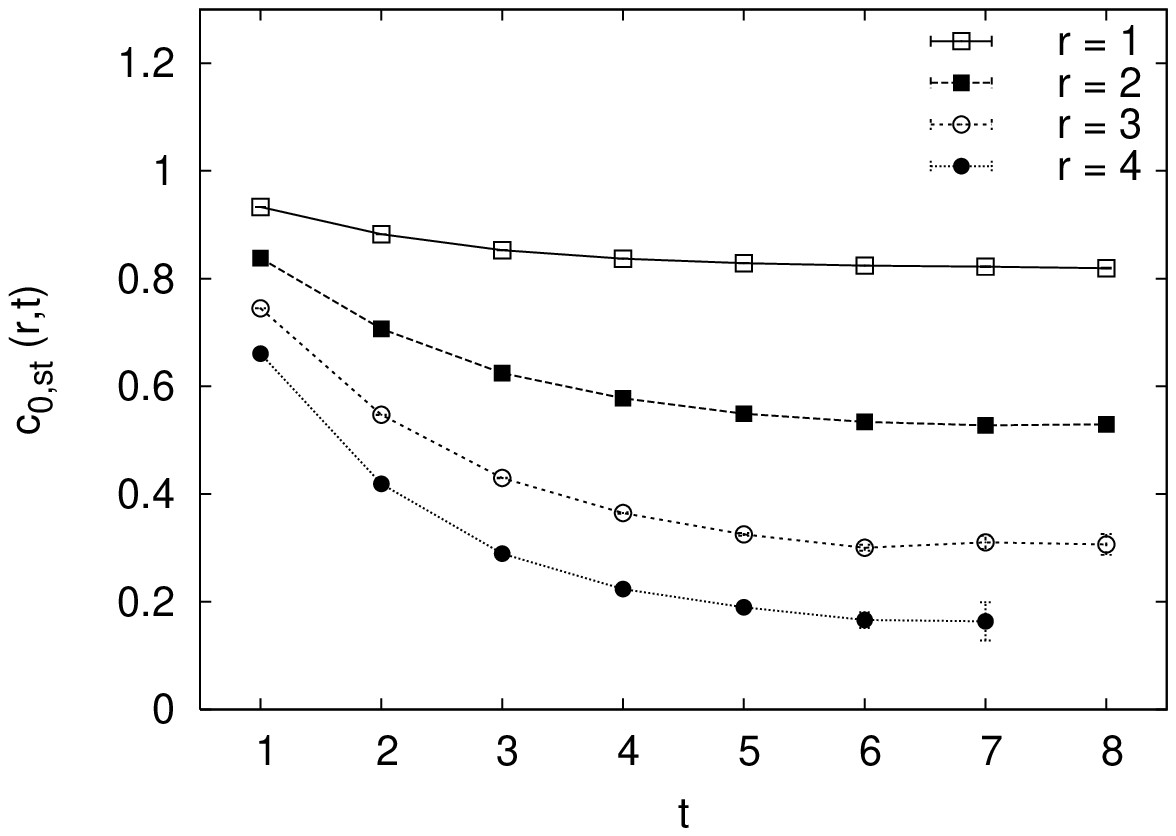} \\
\includegraphics[width=8truecm]{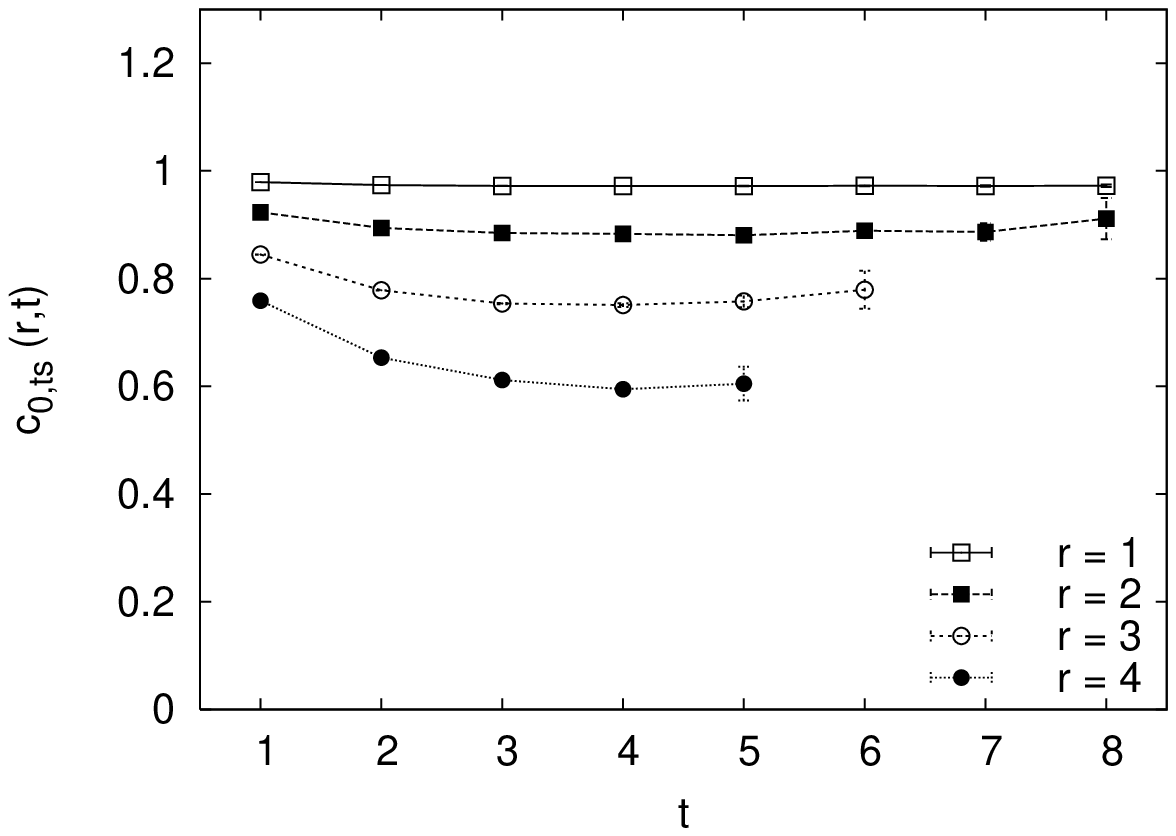} \\
\includegraphics[width=8truecm]{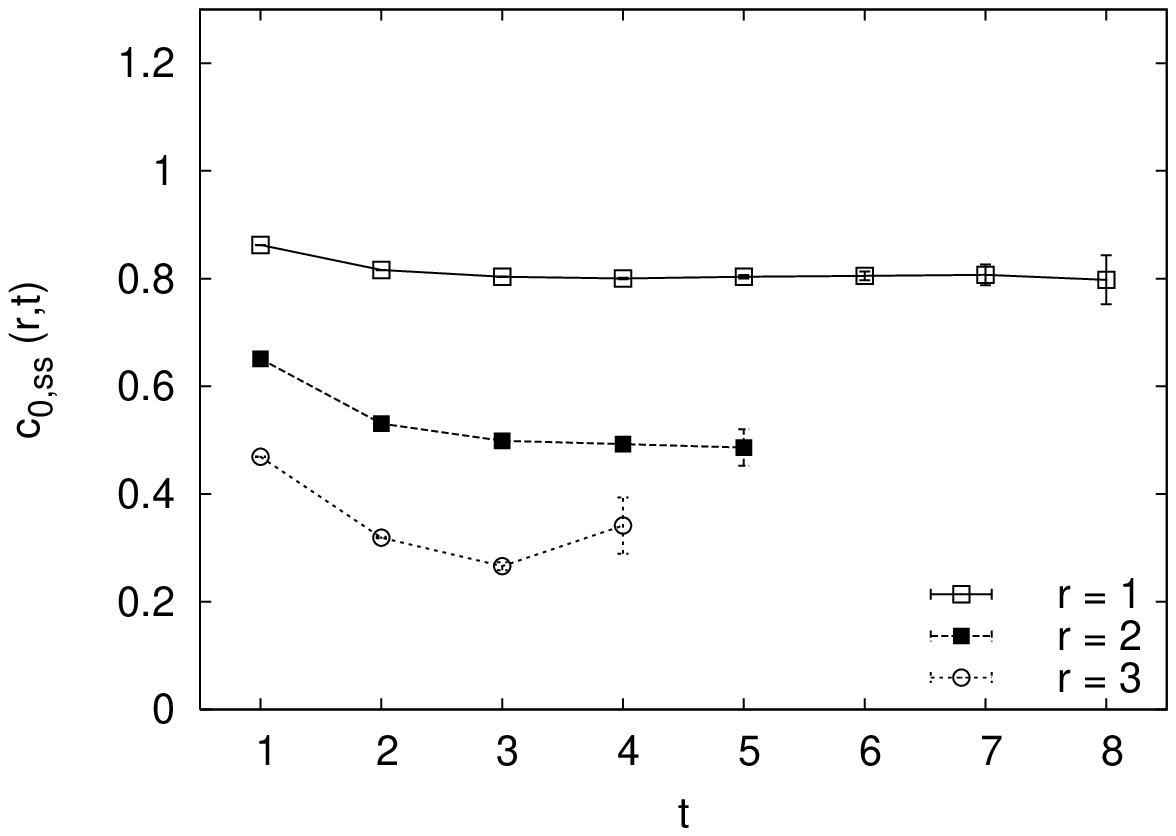} 
\caption{The quantity $c_{0}(r,t)$ as a function of $t$ in the \textit{unsmeared} case for the fundamental representation. ($14^3 \times 28$ lattice, $\beta = 9.6$.)}
\label{fig:ov96usm}
\end{figure}
\begin{figure}[!tbhp]
\includegraphics[width=8truecm]{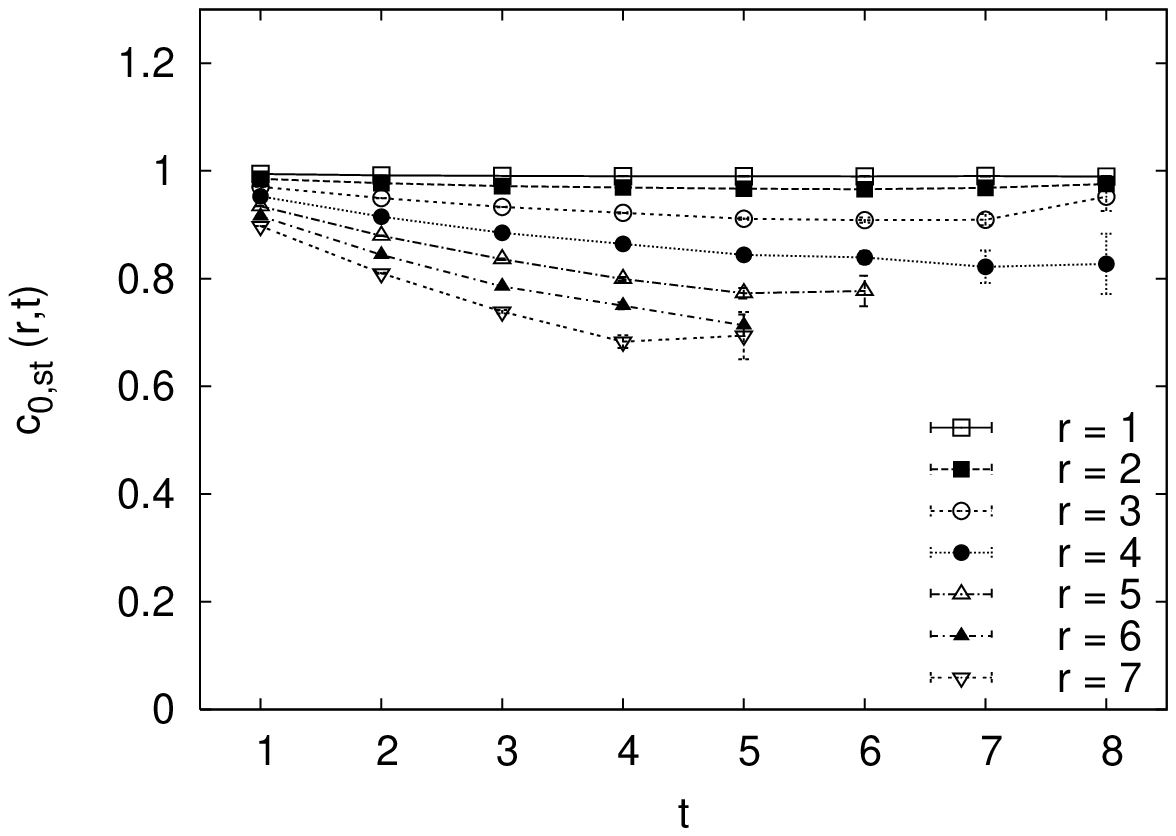} \\
\includegraphics[width=8truecm]{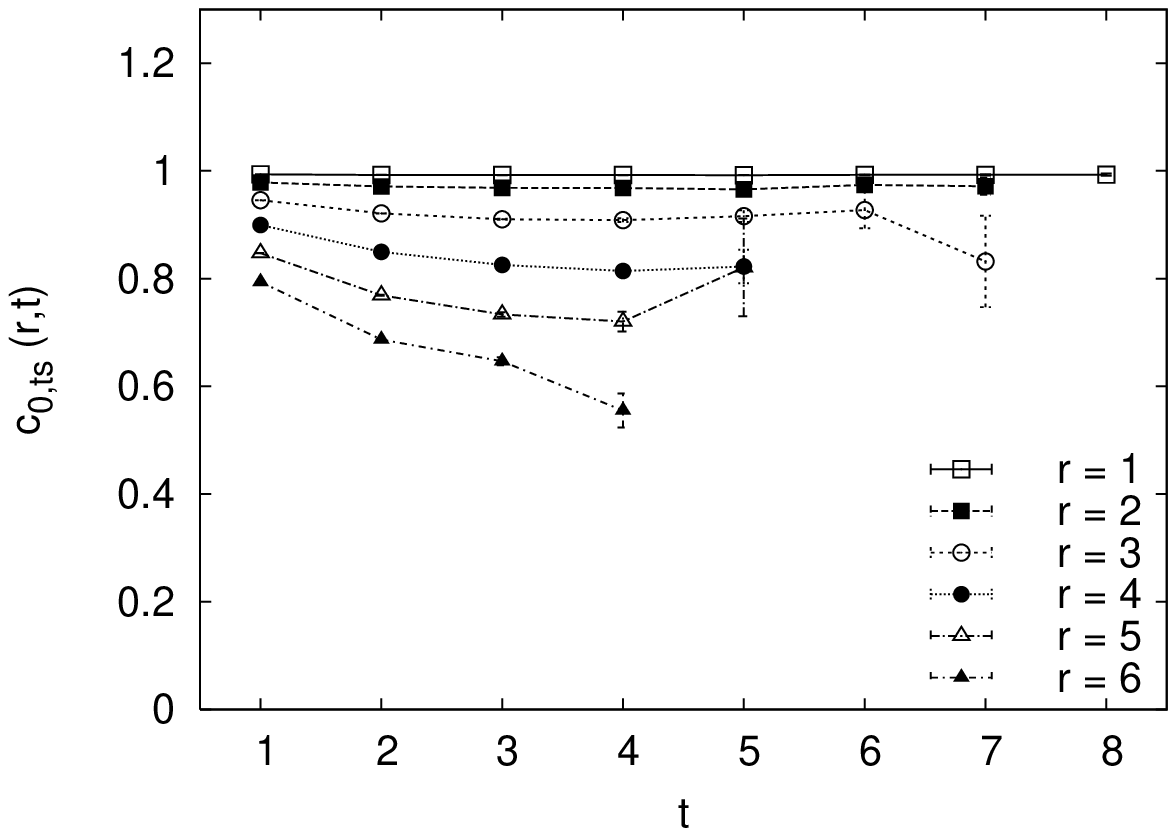} \\
\includegraphics[width=8truecm]{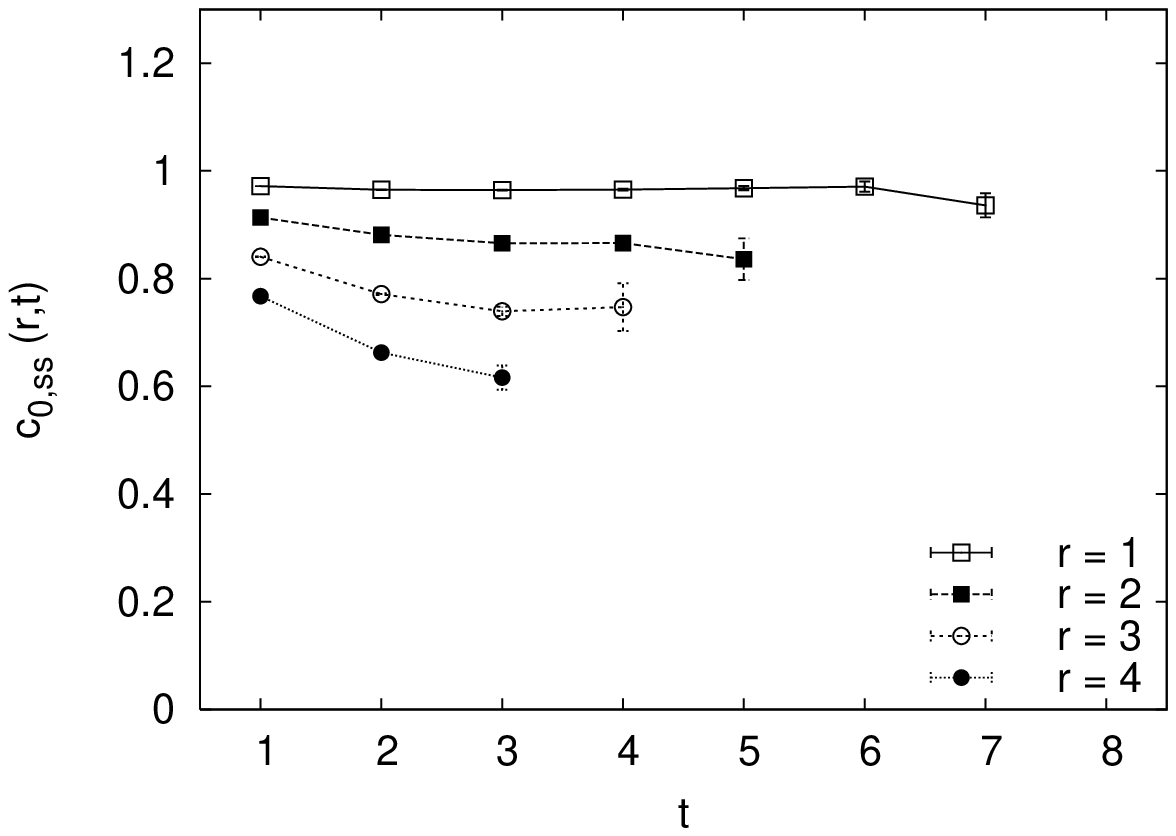}
\caption{The quantity $c_{0} (r,t) $ as a function of $t$ in the \textit{smeared} case for the fundamental representation. ($14^3 \times 28$ lattice, $\beta = 9.6$.)}
\label{fig:ov96sm}
\end{figure}
\begin{figure}[!tbhp]
\centering
\includegraphics[width=8truecm]{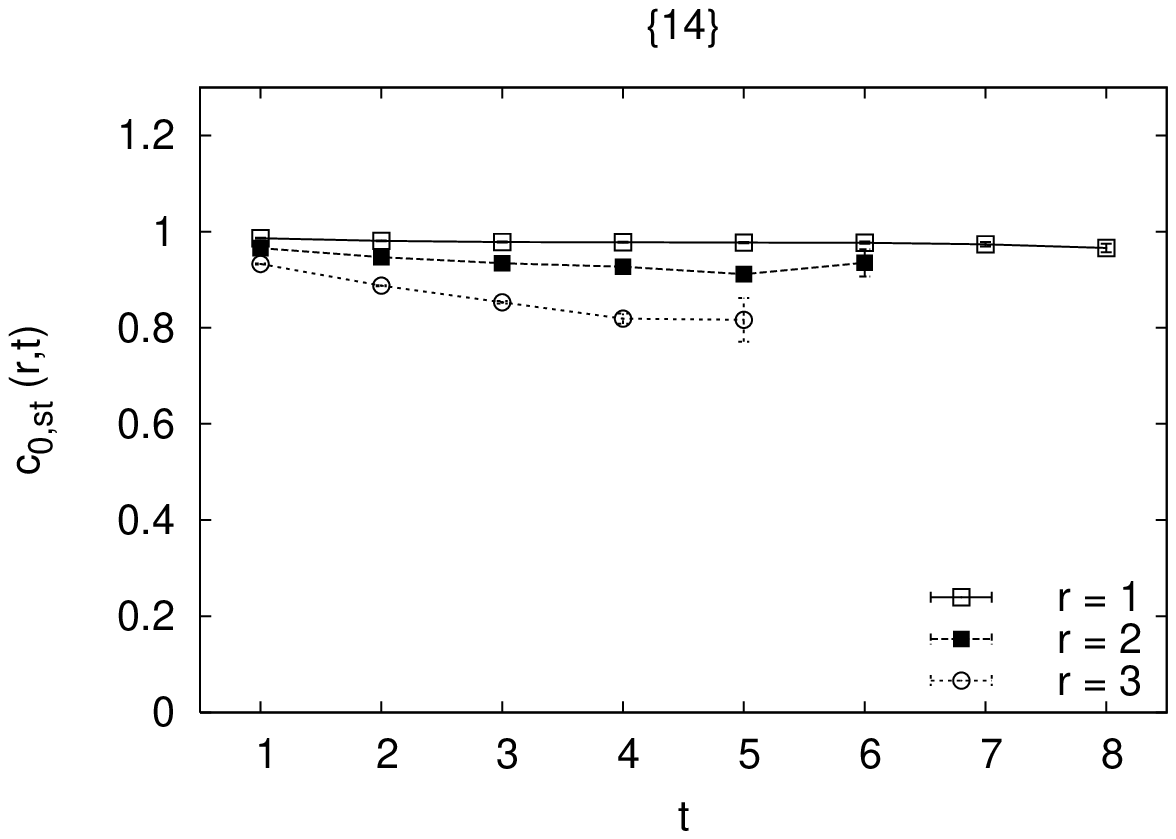} \\
\includegraphics[width=8truecm]{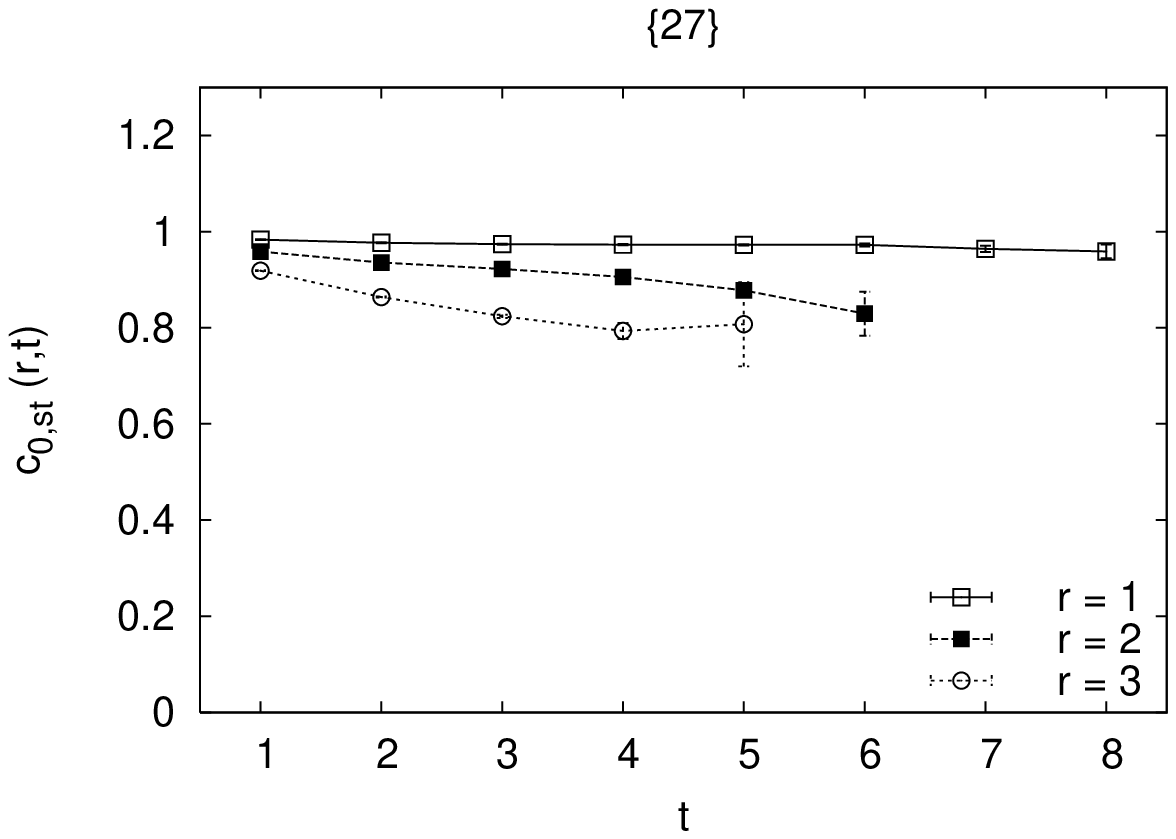} \\
\includegraphics[width=8truecm]{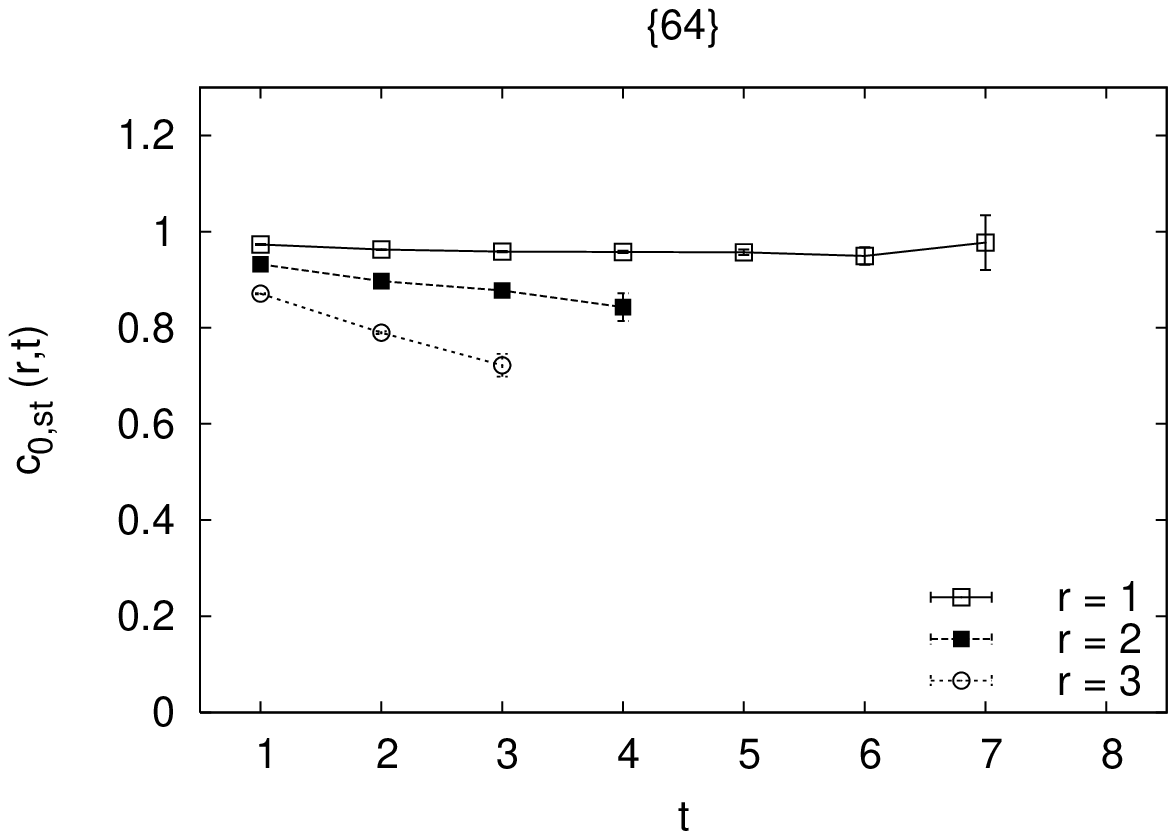}
\caption{The quantity $c_{0,st}(r,t) $ as a function of $t$ in the smeared case for the representations $ \{ 14 \}, \{ 27 \}, \{ 64 \}$. ($14^3 \times 28$ lattice, $\beta = 9.6$.)}
\label{fig:ov96hr}
\end{figure}

	In optimizing our smearing procedure and monitoring the overlap with the ground state, we were inspired by~\cite{Bali:1994de}. We investigate, at fixed $r$, a function of expectation values of Wilson loops defined as:
\begin{equation}
\label{eq:bal1}
c_{0}(r, t)  = \frac{W (r,t)^{t+1}}{W (r,t+1)^{t}} 
\end{equation}
This is an approximant to the ground-state overlap (to be reached in the $t\to\infty$ limit), and decreases monotonically (in $t$) to its asymptotic value. So an indication of the closeness to asymptotia is when the quantity reaches a plateau as a function of $t$.

	Another quantity of interest is:
\begin{equation}
\label{eq:bal2}
M_{\mathrm{eff}}(r, t)  = -\frac{1}{2}\ln\left[\frac{W (r,t+1)}{W (r,t-1)}\right],
\end{equation}
which approaches, from above, the ground-state potential for $t\to\infty$.

	For an asymmetric lattice, we can define three different types of quantities $ c_{0} (r,t) $, corresponding to three different potentials:

\begin{equation}
\label{eq:bal1a}
c_{0,st} (r, t)  =  \frac{  W_{st} (r,t)  ^{t+1} }{  W_{st} (r,t+1)  ^{t} }\ ,
\end{equation} 

\begin{equation}
\label{eq:bal1b}
c_{0,ts} (r, t)  =  \frac{  W_{st} (t,r)  ^{t+1} }{  W_{st} (t+1,r)  ^{t} }\ ,
\end{equation} 

\begin{equation}
\label{eq:bal1c}
c_{0,ss} (r, t)  =  \frac{  W_{ss} (r,t)  ^{t+1} }{  W_{ss} (r,t+1)  ^{t} }\ ,
\end{equation} 
and likewise three types of $M_{\mathrm{eff}}(r,t)$.

	The smearing procedure described in Sec.~\ref{subsec:smearing} depends on the parameter $\rho$ and the number of smearing steps $N_{\mathrm{sm}}$. Our aim was to tune the set to maximize the ground-state overlap and to minimize $M_{\mathrm{eff}}$. Typically, the optimal number of smearing steps grows by decreasing the value of $\rho$. After a few trials we fixed $\rho=0.1$ and looked for the optimal value of $N_{\mathrm{sm}}$. The typical behavior of 
$c_{0,st} (r, t)$ and $M_{\mathrm{eff},st} (r, t)$ for three sets of $(r, t)$ is displayed in Fig.~\ref{fig:klur01}.
In this particular case, the optimal number of smearing steps comes out to be $N_{\mathrm{sm}}\simeq 20$, and depends only weakly
on $\beta$ and $\xi_0$. In a similar way we found the optimal values of $N_{\mathrm{sm}}$ quoted in Sec.~\ref{subsec:smearing}.\footnote{One should keep in mind -- quoting from an extensive study of smearing by D\"urr \cite{Durr:2004xu} -- that ``any smearing recipe drastically reduces the noise and a detailed tuning
of parameters and/or iteration level is neither needed nor useful. [...] Only at absurdly high levels systematic effects get visible, presumably due to the reduced locality in terms of the original links.''}

	As a final illustration, we present here the behavior of the achieved ground-state overlaps as functions of $t$ for $ \beta = 9.6 $. (For $\beta = 9.5$ and $\beta = 9.7$ we observed the same behavior, only errorbars were slightly different.) Results are shown in Figs.\ \ref{fig:ov96usm}, \ref{fig:ov96sm} and \ref{fig:ov96hr}. In all figures only data with reasonably small errors are shown, for the sake of readability. In Fig. \ref{fig:ov96usm} and \ref{fig:ov96sm} the approximants of ground-state overlaps are shown in the unsmeared and smeared cases, respectively, for the fundamental representation: we see clearly that the smearing procedure significantly increases the overlaps and the length of plateaus for all potentials. In Fig. \ref{fig:ov96hr} we display the behavior of  $c_{0,st} (r,t)$ for higher representations and see a similarly good enhancement of the overlap with the ground state. Of course, the overlaps worsen when going to higher representations. 

%

\end{document}